\documentclass[journal]{IEEEtran}

\usepackage[nocompress]{cite}
\usepackage{amsmath,amssymb,amsfonts}
\usepackage{algorithm}

\usepackage{listings}
\usepackage{algorithm}
\usepackage{graphicx}
\usepackage{textcomp}
\usepackage{algpseudocode} 
\usepackage[dvipsnames]{xcolor}
\usepackage{multicol}

\usepackage{enumitem}
\usepackage{tabularx}

\colorlet{shadecolor}{lightgray!40}

\definecolor{red1}{RGB}{218, 65, 61}
\definecolor{gray1}{RGB}{156,169,181}
\definecolor{blue1}{RGB}{68,102,165}

\begin{document}

\markboth{Accepted in IEEE Transactions on Computers}{}
\title{Optimizing Structured-Sparse Matrix Multiplication in RISC-V Vector Processors}

\author{Vasileios Titopoulos, Kosmas Alexandridis, Christodoulos Peltekis,\\ Chrysostomos Nicopoulos, and Giorgos Dimitrakopoulos 
\thanks{
Vasileios Titopoulos, Kosmas Alexandridis, Christodoulos Peltekis, Giorgos Dimitrakopoulos are with the Department
of Electrical and Computer Engineering, Democritus University of Thrace, Xanthi, Greece\protect\\
E-mail: \{vtitopou, koalexan, cpeltekis, dimitrak\}@ee.duth.gr}
\thanks{Chrysostomos Nicopoulos is  with the Department
of Electrical and Computer Engineering, University of Cyprus, Nicosia, Cyprus.\protect\\
E-mail: nicopoulos@ucy.ac.cy}
\thanks{This work was supported by a research grant from Codasip, a provider of customizable RISC-V IP and Codasip Studio design toolset, and its University Program to Democritus Univ. of Thrace for ``RISCV vector processor design''.}}

\maketitle

\begin{abstract}
Structured sparsity has been proposed as an efficient way to prune the complexity of Machine Learning (ML) applications and to simplify the handling of sparse data in hardware. Accelerating ML models, whether for training, or inference, heavily relies on matrix multiplications that can be efficiently executed on vector processors, or custom matrix engines. This work aims to integrate the simplicity of structured sparsity into vector execution to speed up the corresponding matrix multiplications. Initially, the implementation of structured-sparse matrix multiplication using the current RISC-V instruction set vector extension is comprehensively explored. Critical parameters that affect performance, such as the impact of data distribution across the scalar and vector register files, data locality, and the effectiveness of loop unrolling are analyzed both qualitatively and quantitatively. 
Furthermore, it is demonstrated that the addition of a single new instruction would reap even higher performance. The newly proposed instruction is called {\tt vindexmac}, i.e., vector index-multiply-accumulate. It allows for indirect reads from the vector register file and it reduces the number of instructions executed per matrix multiplication iteration, without introducing additional dependencies that would limit loop unrolling. The proposed new instruction was integrated in a decoupled RISC-V vector processor with negligible hardware cost. Experimental results demonstrate the runtime efficiency and the scalability offered by the introduced optimizations and the new instruction for the execution of state-of-the-art Convolutional Neural Networks. More particularly, 
the addition of a custom instruction improves runtime by 25\% and 33\%, 
when compared with highly-optimized vectorized kernels that use only the currently defined RISC-V instructions.
\end{abstract}


\section{Introduction}
\label{sec:introduction}

Machine Learning (ML) applications have been remarkably successful in various application domains. To reduce memory storage and computation cost, the weights of ML models are often pruned, thereby leading to sparse models~\cite{hoefler2021sparsity}. This means that zero weights are not stored, and the corresponding calculations are skipped. While sparsity offers high performance and low storage overhead, it may, in some cases, result in less accurate ML models~\cite{sparse-tensor-core, hoefler2021sparsity}. In such cases, accuracy loss can be mitigated by retraining the model to adapt to the removal of specific weights, as demonstrated for CNNs~\cite{learning-n-m,nvidia-block-sparse} and Transformers~\cite{elsa,jiang2022exposing}.

\begin{figure}[t]
\centering
\includegraphics[width=0.95\columnwidth]{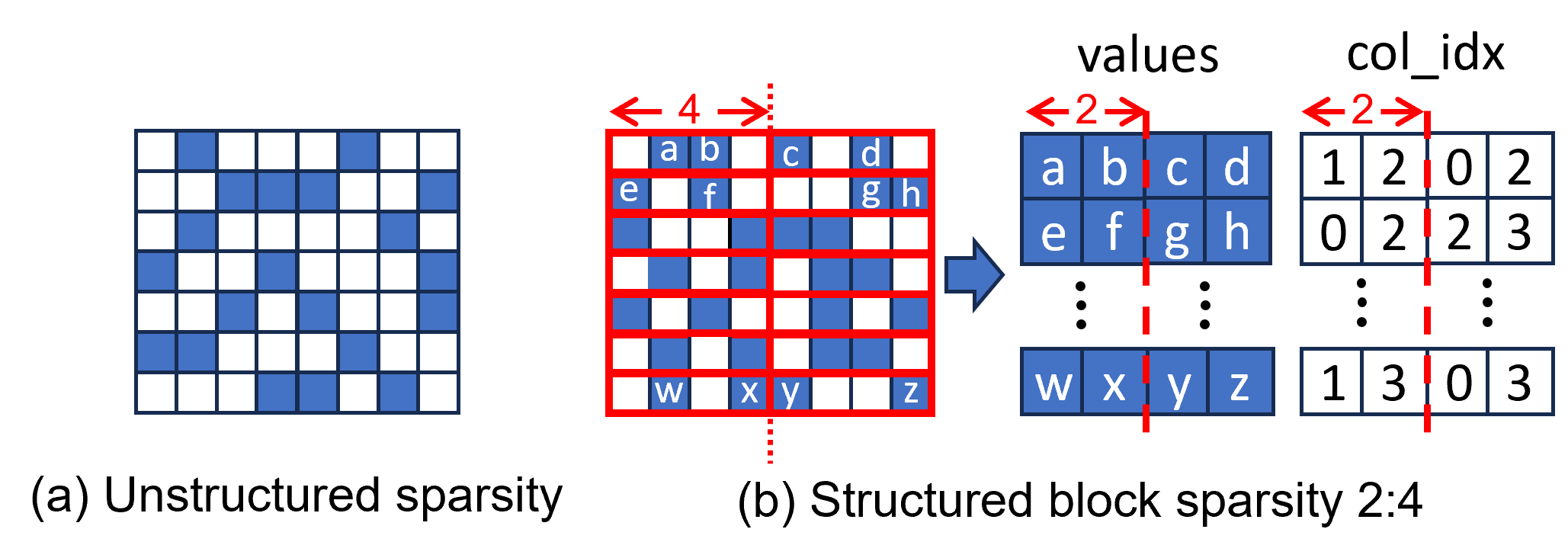}
\caption{Example of (a) unstructured sparsity; and (b) structured block sparsity of 2:4 (i.e., up to 2 non-zero elements in every 4 consecutive elements) and their respective representation. Blue squares represent non-zero elements.}
\label{f:unstructered-block-sparse}
\end{figure}

The achieved sparsity can either be \textit{unstructured}~\cite{rigl}, or \textit{structured}~\cite{nvidia-block-sparse,learning-n-m}. In unstructured sparsity, there is no constraint on the locations of the zeros, as shown in Fig.~\ref{f:unstructered-block-sparse}(a). In this case, together with the non-zero elements, multiple indexes are also required to identify the original position of each non-zero element. 

On the contrary, in structured sparsity, there is an upper limit on the number of non-zero elements that may be present within a block of consecutive elements. For instance, in Fig.~\ref{f:unstructered-block-sparse}(b), for every 4 elements in each row, there are up to two non-zero elements. In most practical applications~\cite{nvidia-block-sparse,s2ta}, blocks are small and $N$:$M$ sparsity patterns of 1:2, 1:4 or 2:4 are supported, where each block of $M$ elements may contain up to $N$ non-zero elements. As also shown in  Fig.~\ref{f:unstructered-block-sparse}(b), for such block-based structures, the indexing ({\tt col\_idx}) required to identify the position of each non-zero element inside each block consists of just of few bits (owing to the small size $M$ of each block). Recent work has shown that matrices with unstructured sparsity can be transformed offline to structured-sparse matrices either by row (column) reordering and zero padding~\cite{polyomino} or matrix splitting~\cite{split-to-structured-sparse}. 
Tensor units on GPUs can accelerate ML applications with structured sparse data~\cite{learning-n-m,nvidia-block-sparse}. Our goal is to extend this hardware capability to CPUs with vector engines, making it applicable to a broader range of current and future applications.

Un-structured sparsity is already supported in vector processors.
Various sparse vector matrix multiplication algorithms have been developed~\cite{spa, esc}, and recently extended~\cite{generic-spgemm, upc-spgemm}, with the goal being to improve performance by efficiently handling unpredictable sparsity patterns. Moving one step further, VIA~\cite{via} adds a scratchpad memory and new custom vector instructions to deal with the complexity of unstructured sparse data.

Aiming to avoid the substantial hardware overhead incurred in exploiting \textit{unstructured} sparsity, this work similar to~\cite{islped} makes the case that \emph{structured} sparsity, which appears frequently in state-of-the-art CNN applications~\cite{nvidia-block-sparse}, can be exploited very effectively -- and with negligible hardware cost -- in vector processors to achieve high-performance sparse$\times$dense matrix ($A\times B$) multiplications. 

First, through extensive analysis, we demonstrate how to reorganize and optimize the state-of-the-art row-based matrix multiplication algorithm~\cite{generic-spgemm, matraptor, arm-patent} for structured-sparse data in long-vector Instruction Set Architectures (ISAs), such as RISC-V. To better adjust to the available vector multiply-and-accumulate instructions of the RISC-V ISA vector extension, a mixed placement of non-zero data and their corresponding column indexes across the vector and the scalar register files is proposed, which reduces register name dependencies and simplifies loop unrolling. 

Secondly, a new \textit{custom} vector-index-multiply-accumulate instruction is proposed, called {\tt vindexmac},
which can lead to substantial \textit{further} performance improvement. 
This instruction enables the implementation of low-cost indirect reads from the vector register file that eliminates unnecessary memory traffic often encountered in sparse matrix multiplication algorithms.  
The proposed vector instruction was integrated in a high-end decoupled RISC-V vector processor at negligible hardware cost. Decoupled vector processors~\cite{ara, riscv-vector, vitruvius,tarantula} have independent decode and execute pipelines, separate from the scalar core, which fetches and forwards the corresponding vector instructions. Additionally, a decoupled vector processor accesses memory (or the cache hierarchy) independently of the scalar core, unlike integrated vector engines~\cite{integrvec} that share the memory access path with the scalar core. 

The presented evaluation employs the gem5 simulator~\cite{gem5-orig, gem5-2020} implementing an accurate model of a decoupled vector processor attached to a superscalar out-of-order core. The executed benchmark applications comprise state-of-the-art Convolutional Neural Networks (CNNs) pruned for structured sparsity. The obtained results demonstrate: (a) the significant speedup that can be achieved when optimizing data placement across the register files of a RISC-V-compliant vectorized kernel of structured-sparse matrix multiplication; (b) the additional speedup that can be achieved when the vector ISA is extended with the new {\tt vindexmac} custom multiply-index-accumulate vector instruction. 

Additional experiments show that performance scales favorably with increasing hardware vector length, or by splitting CNN layer execution across multiple vector processors in a multicore setup. However, in the latter case, the benefits saturate beyond 8 cores, due to memory bandwidth limitations.

The contributions of this work can be summarized as follows:
\begin{itemize}
\item 
Mixed data placement of structured-sparse data across vector and scalar register files reduces register name dependencies and allows for loop unrolling optimization for sparse matrix multiplication in vector processors. The presented design space exploration identifies which configuration yields the lowest execution time under the current definition of the RISC-V ISA.
\item 
A new vector index, multiply, and accumulate instruction enables multiply and accumulate to be performed on vectors read from the vector register file, both directly and indirectly. This instruction can be seamlessly integrated into a vector processor at negligible cost. Its use promotes data locality in the vector register file and reduces the number of instructions per iteration.
\item 
The proposed optimizations and custom instruction significantly improve the runtime efficiency and scalability of state-of-the-art Convolutional Neural Networks. Specifically, the custom instruction leads to a 25\%--33\% runtime improvement over highly-optimized vectorized kernels using only existing RISC-V instructions. 
\end{itemize}

The rest of the paper is organized as follows: Section~\ref{s:vectorized_sparse} presents the formulation of a state-of-the-art vectorized algorithm for sparse matrix multiplications with structured-sparse data and its optimized variants. Section~\ref{s:optimization} introduces the proposed vector index-multiply-accumulate instruction and its usage in the structured-sparse matrix multiplication kernel. Experimental results are presented in Section~\ref{s:eval}, and conclusions are drawn in Section~\ref{s:conclusions}.

\section{Design Space Exploration of Vector Sparse$\times$Dense Matrix Multiplication }
\label{s:vectorized_sparse}

\begin{figure}[t]
\centering
\includegraphics[width=0.8\columnwidth]{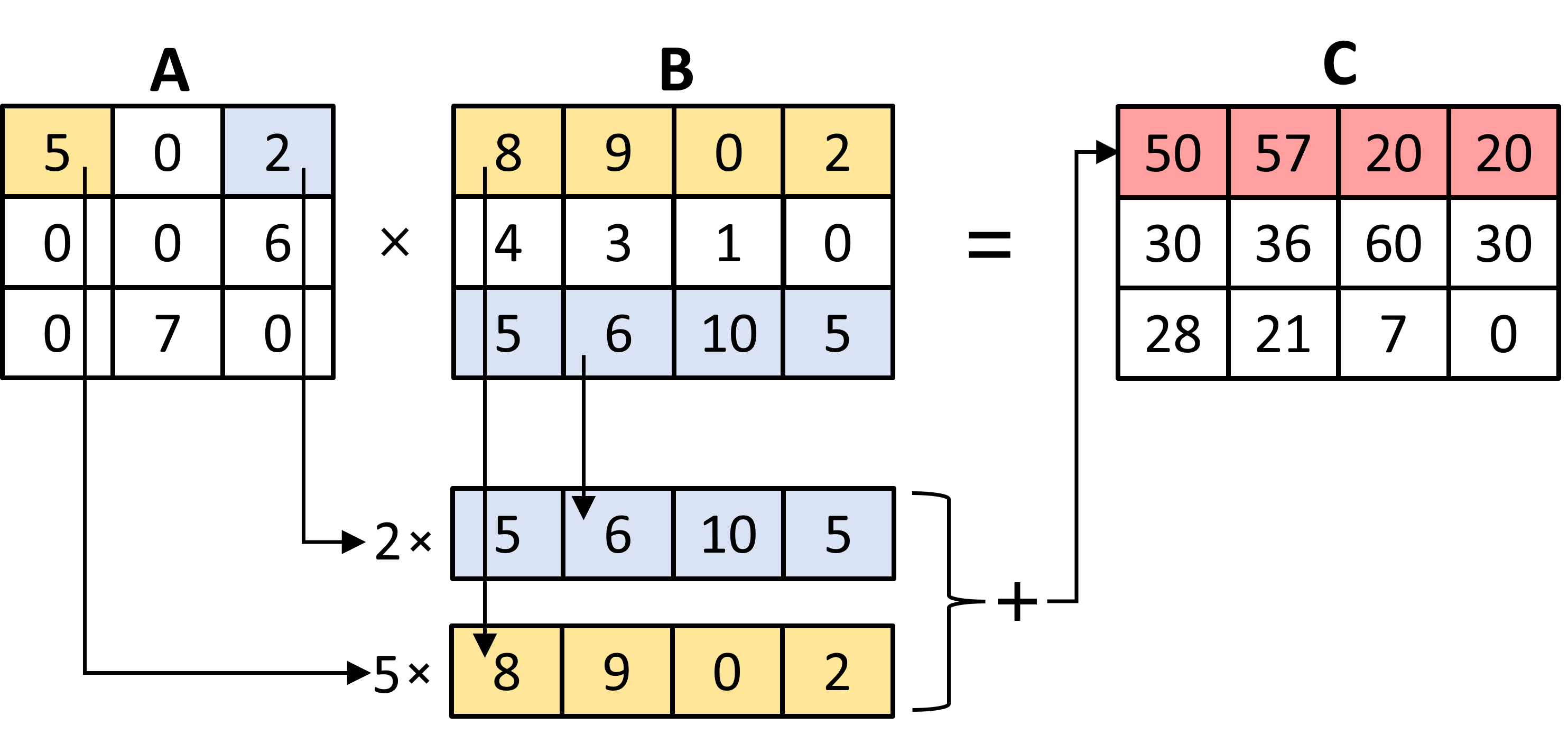}
\caption{Row-wise matrix multiplication to compute output row $C[0,:]$.} \label{f:row-wise_multiplication}
\end{figure}

Vectorized matrix multiplications with sparse data can be implemented with many approaches~\cite{spa, generic-spgemm, upc-spgemm}. The row-wise approach~\cite{matraptor, mm-gp-simd, takayashiki2023new}, also known as Gustavson’s algorithm~\cite{gustavson}, has been shown to be highly effective in computing the matrix product $A\times B$, and it is a better fit to the targeted structured sparsity context. Other approaches~\cite{spa,generic-spgemm,upc-spgemm} target extremely high sparseness and are not as effective with modest structured sparsity.

Matrix $A$ is assumed to be structured sparse and matrix $B$ is considered to be dense. Algorithmically, all the non-zero elements in a single row of matrix $A$ should be multiplied in parallel with the corresponding rows of matrix $B$, where the row index of matrix $B$ is determined by the column index of the non-zero value in matrix $A$. The product of the multiplication is produced row-by-row, as follows: 
\begin{equation}
C[i,:] = \sum_{k} A[i,k] B[k,:]
\end{equation}
Fig.~\ref{f:row-wise_multiplication} illustrates a simple example of how row $0$ of the result matrix $C$ is produced. Row-wise matrix multiplication can be easily vectorized, since each element of $A$ is multiplied with \textit{all} the elements of a row of matrix $B$. This yields a vector of partial results for a row in the result matrix $C$.

\begin{figure*}[t!]
\centering
\includegraphics[width=2\columnwidth]{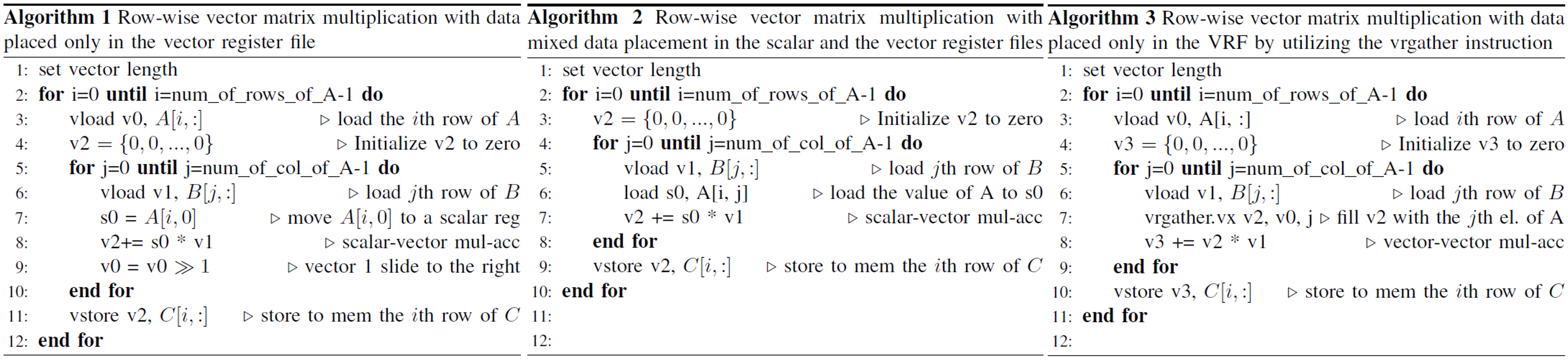}
\label{f:trial}
\end{figure*}

\subsection{Exploring data placement in vector row-wise matrix multiplication}

The vectorized version of the row-wise matrix multiplication can take three equivalent forms, albeit with different performance characteristics. The first approach, depicted in Algorithm 1, assumes that the elements of matrices $A$ and $B$ are placed exclusively in the Vector Register File (VRF). Algorithm 1 \emph{ignores, for clarity,} any sparseness in matrix $A$ and assumes an arbitrarily large vector length. 
In line 3, all the elements of row $i$ of matrix $A$ are loaded as a vector into the VRF. Similarly, all the elements of the corresponding row of matrix $B$ are also loaded as a vector in line 6.  
The vector multiplication between the first element of the loaded row of $A$ and the entire row of matrix $B$ is performed in line 8, which also accumulates the partial results. Following the RISC-V vector ISA format, the multiply-add operation of line 8 is performed on a vector register and a scalar value coming from the scalar register of line 7. The row of $A$ is shifted to the right by one element in line 9 to enable the repetition of the above process for the remaining elements of the same row. The final values of all the elements of the corresponding row of the result matrix $C$ are stored back in memory in line 11. The process is repeated until all the rows of matrix $C$ are produced.

However, the placement of the elements of matrix $A$ in the VRF in Algorithm 1 features many unnecessary data transfers between the VRF and the scalar RF (line 7). To tackle this inefficiency, we explore two alternative approaches: (a) keep the elements of matrix $B$ in the VRF and place the elements of matrix $A$ entirely in the scalar RF; and (b) utilize more vector registers in the VRF for the non-zero elements of matrix $A$.

Starting with the first approach, Algorithm 2 presents the row-wise vector matrix multiplication that follows the above-mentioned \textit{hybrid} data placement. Specifically, rather than loading the rows of matrix $A$ into the VRF, each element of a row of $A$ is, instead, loaded (line 6) and used (line 7) one-by-one directly through the scalar RF. Thus, the multiply-add operation in line 7 directly utilizes the data placed in the scalar register `S0,' thereby eliminating the need for vector-to-scalar register moves and vector slides. Note that the data type of the elements dictates the type of the register file into which the non-zero values will be stored: if the non-zero values in Algorithm 1 and Algorithm 2 are of type `float,' they can be stored into the floating-point register file, thereby alleviating the overall pressure on the scalar register file.

Alternatively, the second approach is to keep matrix $A$ in the VRF and utilize more vector registers, as shown in Algorithm 3. Line 3 shows that a row of matrix $A$ is loaded into the VRF. We need to access each one of those elements one after the other and multiply them with the corresponding row of $B$. This is achieved using the {\tt vrgather.vx} instruction. Said instruction copies the $j$th element of `v0' into every element of the vector register `v2.' The latter vector register is then used as an operand in the multiply and accumulate operation. The use of the {\tt vrgather.vx} instruction enables the operation of the multiply-and-accumulate operation (line 8) to be performed with 3 vector operands, as opposed to the approach in Algorithm 2, where a scalar and 2 vector operands are used. 

\begin{figure*}[t]
\centering
\includegraphics[width=2\columnwidth]{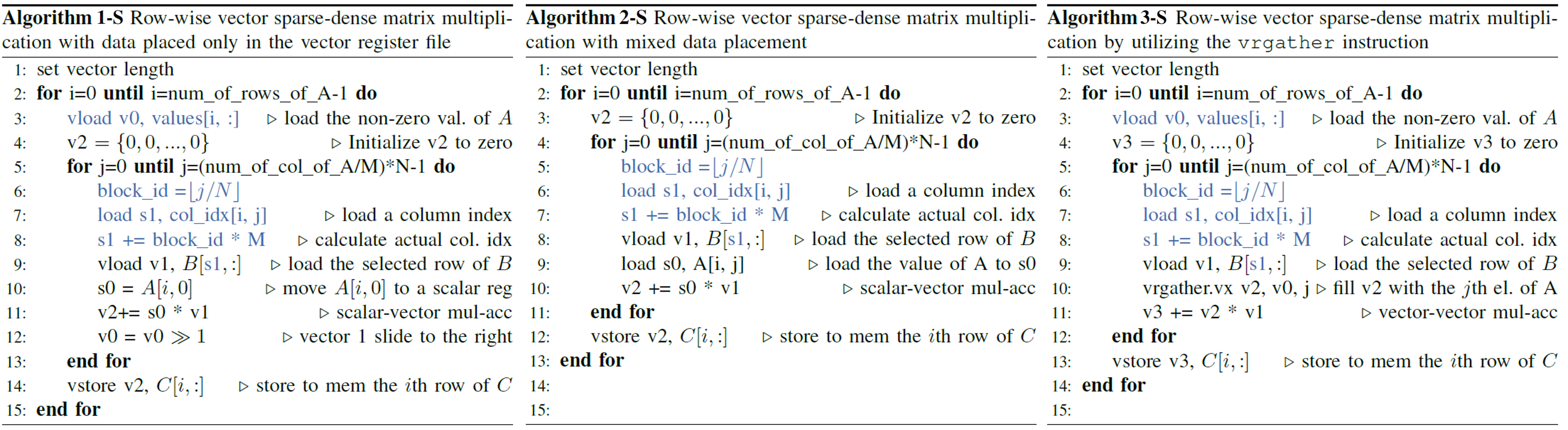}
\label{f:trial2}
\end{figure*}

\subsection{Vector Row-wise matrix multiplication for structured-sparse data}

Algorithms 1--3 can be reformulated to support a \textit{structured-sparse} matrix $A$. In this case, 
only the non-zero elements of each row of $A$ are stored, together with their corresponding index, i.e., their original position within a block of size $M$ in the row of matrix $A$.
The reformulation of Algorithms 1-3 is shown in Algorithms 1-S, 2-S and 3-S respectively, where all mandatory changes/additions are highlighted with color.

To highlight the added changes we will use as an example Alg. 3--S. In this case, 
the non-zero values of matrix $A$ are loaded into the VRF, as shown in line 3 of Algorithm 3-S. 
On the contrary, the column indexes are stored in a \textit{scalar} register in the scalar RF (as per line 7 in the algorithm), because they require some pre-processing before they can be used. The pre-processing steps arise from the structured sparsity format itself, which limits the range of the column indexes. Specifically, the column index refers to the position of the non-zero element inside a block of size $M$. Consequently, all column indexes have a range $[0, M-1]$, even though each row of matrix $A$ typically consists of \textit{multiple} blocks of size $M$.

\begin{figure}[htb]
\centering
\includegraphics[width=0.9\columnwidth]{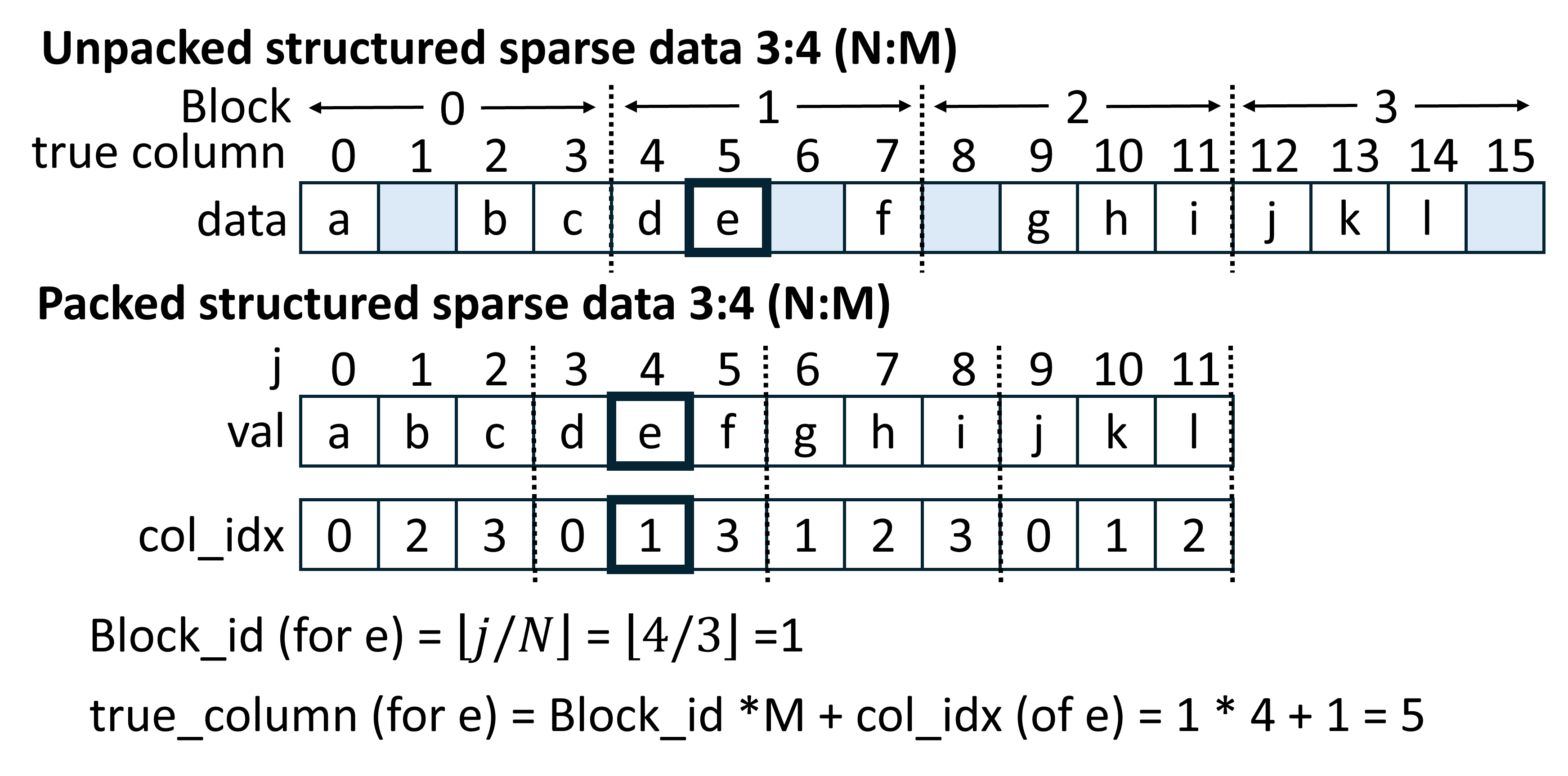}
\caption{An example of recovering the actual column index of an element through the column index that is stored in the memory and the block\_id.} 
\label{f:actual col idx}
\end{figure}

Hence, whereas the column index `j' in Algorithm 3 spans the entire row of matrix $A$ (i.e., all its columns), the same index has a different meaning in Algorithm 3-S. Column index `j' now spans only the number of non-zero elements in a particular row of matrix $A$, as shown in line 5 of Algorithm 3-S. In this form, the index cannot be used directly to load the corresponding row of matrix $B$. Instead, the actual column index that points to a specific column across the \textit{entire} row of matrix $A$ must first be retrieved. To compute the actual column index, we must identify the particular block in which the corresponding non-zero element resides. This is performed in line 6 of the algorithm, where the block index is calculated by the integer division of the position `j' of the non-zero element with the maximum permissible number of non-zero elements $N$ in each block of size $M$.  
Once the block ID is identified, it is multiplied with the size $M$ of each block and added to the original column index that was loaded from memory, as shown in line 8. This operation fully retrieves the actual column index that spans the entire row of matrix $A$. When $N$ and $M$ are multiples of 2, multiplication and division are performed with left and right shifts respectively.

An example of how the actual column index can be retrieved is depicted in Fig.~\ref{f:actual col idx}, assuming 3:4 ($N$:$M$) structured block sparsity. The example focuses on element `e.' First, the block in which element `e' resides (block\_id) is identified by performing an integer division between the position of `e' in the array of non-zero values (i.e., position 4), and the number $N$=3 of permitted non-zero elements in each block of size $M$. Next, the actual column index is computed by multiplying the block\_id with the size $M$=4 of each block. Finally, this value is added to the column index of element `e' that is stored in memory (i.e., the value 1) to retrieve the actual column within the row of $A$ where element `e' resided prior to compression.

To avoid such indexing calculations, full-width column indexes can be stored at the row level. However, as demonstrated in~\cite{nvidia-block-sparse} and in Section~\ref{ss:current-ISA} this approach would significantly increase storage overhead 
for the targeted sparsity levels of $1$:$4$ or $2$:$4$.

Similarly, the low-to-medium levels of sparsity in the context of a sparse-dense multiplication, which is the focus of this work, make it less preferable to gather the non-zero values of \( A \) using indexed load instructions~\cite{upc-spgemm}. This approach would introduce additional runtime overhead, since indexed vector load/store instructions may generate a substantial number of cache misses from only a few vector instructions. 

Continuing with Algorithm 3-S, the actual column index retrieved in line 8 is used to load the corresponding row of matrix $B$ into the VRF in line 9. The {\tt vrgather} instruction is used in line 10 to copy the currently-processed non-zero element into all the vector register elements of vector `v2.' Note that only the rows of matrix $B$ that correspond to the column indexes of the non-zero elements of $A$ participate in the multiply-and-accumulate operation (line 11). Each iteration of the inner loop moves to the next \textit{non-zero} element in the current row of $A$, until the end of the row is reached. 

Algorithms 1-S and 2-S employ a similar approach to retrieve the non-zero values of A and their corresponding column indices. Section~\ref{ss:current-ISA} highlights the relative performance of Algorithms 1-S, 2-S, and 3-S, focusing on optimized unrolled implementations. The following section demonstrates how loop unrolling is optimized for Algorithm 3-S, which benefits most from this optimization.

\subsection{Loop unrolling row-wise vector sparse$\times$dense matrix multiplication}
\label{ss:loop-unrolling}

Following the practice explored in other vector-based parallel matrix multiplication kernels~\cite{arm-sve-mm, micro-kernels}, and to increase the scope for instruction-level parallelism, both loops (outer and inner) of Algorithm 3-S can be unrolled. 
Indicatively, Algorithm~\ref{a:vectorized_row-wise_sparse-dense-scalar-unrolled-vrgather} unrolls both loops by a factor of 2. The unrolling of the outer loop allows the processor to simultaneously work on \textit{two consecutive} rows of matrix $A$, i.e., to simultaneously produce two consecutive rows of output matrix $C$. Similarly, the unrolling of the inner loop allows the processor to simultaneously work with two consecutive non-zero elements in each of the two consecutive rows of matrix $A$. 

Lines 8 to 18 in Algorithm~\ref{a:vectorized_row-wise_sparse-dense-scalar-unrolled-vrgather} refer to operations performed using the non-zero element 'j' of consecutive rows `i' and `i+1' of matrix $A$. As can be seen, due to the loop unfolding, each instruction (except the one calculating the block\_id) in the inner loop of Algorithm 3-S is hereby duplicated. The two multiply-accumulate instructions in lines 17 and 18 generate two vectors of partial results for rows `i' and `i+1' of the output matrix $C$. Similarly, lines 19 through 29 repeat the same pattern for the next non-zero elements in the same two rows of matrix $A$; i.e., the non-zero element `j+1' of consecutive rows `i' and `i+1.'

Since each loop was unrolled by a factor of 2, one would expect the final impact on the required scalar and vector registers to be four-fold. However, this cost is avoided through the \textit{interleaving} of the two loops, which allows for the reuse of the same scalar and vector registers and reduces read-after-write dependencies.
As can be seen in Algorithm~\ref{a:vectorized_row-wise_sparse-dense-scalar-unrolled-vrgather}, instructions from different loop iterations are executed in interleaved manner. The instructions that belong to the same loop are depicted with the same color. In all cases, no two instructions of the same loop are executed one after the other. 
Interleaving allows for exactly the same scalar and vector registers used in lines 8 to 18 to be reused in lines 19 to 29. This approach minimizes the amount of registers in use, while also maintaining the physical parallelism that is inherent to non-dependent instructions. Also, interleaving allows distinct loop iterations to share the  column-index calculations. This effectively reduces the number of instructions after unrolling.

Overall, Algorithm 3-S requires 1 scalar and 4 vector registers, whereas the loop-unfolded Algorithm~\ref{a:vectorized_row-wise_sparse-dense-scalar-unrolled-vrgather} requires 2 scalar and 8 vector registers. In general, (a) increased pressure on the register file, and (b) code bloating (that could adversely affect the cache performance) are the two well-known side effects of loop unrolling. Hence, loop unrolling should be applied judiciously up to the point where the side effects of the unrolling start to cause diminishing returns to the reaped performance improvement.

\begin{algorithm}[t]
\caption{Unrolled row-wise vector sparse-dense matrix multiplication by utilizing the {\tt vrgather.vx} instruction}
\label{a:vectorized_row-wise_sparse-dense-scalar-unrolled-vrgather}
\begin{algorithmic}[1]
\State set vector length
\For{i=0 \textbf{until} i=num\_of\_rows\_of\_A-1  \textbf{,} \ i+=2}
\State {vload v0, values[i, :]} 
\State {vload v1, values[i+1, :]} 
\State v6 $=\{0,0,...,0\}$
\State v7 $=\{0,0,...,0\}$
\For{j=0 \textbf{until} j=(num\_of\_col\_of\_A/M)*N-1  \textbf{,} \ j+=2}
        \State {block\_id =$\lfloor j/N \rfloor$}   
        \State {\textcolor{blue1}{load s0, col\_idx[i, j]}}  
        \State {\textcolor{gray1} {load s1, col\_idx[i+1, j]}}    
        \State {\textcolor{blue1}{s0 += block\_id * M}}   
        \State {\textcolor{gray1} {s1 += block\_id * M}}   
        \State  \textcolor{blue1}{vload v4, $B[s0, :]$} 
        \State \textcolor{gray1} {vload v5, $B[s1, :]$} 
        \State {\textcolor{blue1}{vrgather.vx v2, v0, j}} 
        \State {\textcolor{gray1}{vrgather.vx v3, v1, j}} 
        \State  \textcolor{blue1}{v6 += v2 * v4} 
        \State \textcolor{gray1} {v7 += v3 * v5} 
        \State {block\_id =$\lfloor (j+1)/N \rfloor$}   
        \State {\textcolor{red1}{load s0, col\_idx[i, j+1]}}  
        \State {\textcolor{Tan}{load s1, col\_idx[i+1, j+1]}}  
        \State {\textcolor{red1}{s0 += block\_id * M}}   
        \State {\textcolor{Tan}{s1 += block\_id * M}}   
        \State  \textcolor{red1}{vload v4, $B[s0, :]$} 
        \State \textcolor{Tan} {vload v5, $B[s1, :]$} 
        \State {\textcolor{red1}{vrgather.vx v2, v0, j+1}} 
        \State {\textcolor{Tan}{vrgather.vx v3, v1, j+1}} 
        \State  \textcolor{Tan}{v6 += v2 * v4} 
        \State \textcolor{red1} {v7 += v3 * v5} 
    \EndFor
  \State vstore v6, $C[i, :]$ 
  \State vstore v7, $C[i+1, :]$ 
\EndFor
\end{algorithmic} 
\end{algorithm}

\section{Speeding up sparse-dense matrix multiplication with a custom instruction}
\label{s:optimization}

The implementation of sparse-dense multiplication eliminates unnecessary multiplications, due to the structured-sparse format of $A$. However, one crucial bottleneck of the computation is the abundance of \textit{vector loads} from memory for the rows of matrix $B$ (Line 9 in Algorithm 3-S). 
To tackle this issue, we leverage the structured sparsity of matrix $A$ to reduce memory traffic and allow the computations to use \textit{local} data that already resides in the vector register file.

The proposed optimization combines: (a) loading rows of matrix $B$ in tiles in the vector register file;  
and (b) a custom index-multiply-accumulate instruction that replaces the vector loads of matrix $B$ with indirect reads of the vector register file. 

The key attribute that enables the loading of matrix $B$ in the register file in groups of multiple rows (instead of the row-by-row approach) is the well-defined, regular structure in the format of matrix $A$. Since the sparsity of $A$ is -- by construction -- \textit{structured}, the blocks within said matrix have a constant and known size. In turn, this implies that the column indexes of the non-zero element values in $A$ are `bounded' by the block size $M$, i.e., all {\tt col\_idx} values reside within the range $[0, M-1]$. Recall that the block size $M$ is the number of consecutive elements within a row of $A$ that can contain up to a specific number ($N$) of non-zero elements. Exploiting this trait of structured sparsity, we may pre-load as many rows of matrix $B$ as our vector register file can accommodate (with some restrictions, as will be explained in Section~\ref{ss:multiplication_with_vindexmac}) and be sure that the column indexes in matrix $A$ will only point to those local rows. On the contrary, with unstructured sparsity, the column indexes are, essentially, `unbounded' and could point to any row of $B$ (thereby rendering the pre-loading of specific rows of $B$ in the register file futile).

The tile size of matrix $B$ that is pre-loaded in the vector register file is $L\times Vector\_Length$; i.e., $L$ rows, with each row having $Vector\_Length$ columns. The number of rows $L$ must be a multiple of $M$.

\subsection{The proposed vector index-multiply-accumulate {\tt\bf vindexmac} instruction}

Since multiple rows of matrix $B$ are now loaded in the VRF, there is a need to \textit{index} into the register file to access a specific row. This is achieved by the proposed new instruction that effectively performs a scalar-vector multiply-and-accumulate operation, in combination with the indexing in the VRF. The scalar value refers to a non-zero element of a row of matrix $A$ and the vector refers to a row of matrix $B$ that is pre-loaded in the VRF and indirectly read by the provided column index.

Overall, the new instruction aims to alleviate the sparse-dense matrix multiplication process from the continuous vector loads of the rows of matrix $B$. Instead, the custom instruction reads the corresponding pre-loaded rows of matrix $B$ directly from the vector register file and operates on them. With this objective in mind, the new {\tt vindexmac} instruction is defined as follows:

{\tt \colorbox{lightgray!40}{vindexmac.vx vd, vs2, rs}}

  {\tt \hspace{0.3cm} vd[i] += vs2[0] * vrf[rs|$_{\texttt{4:0}}$|][i]}

\begin{figure}[t]
\includegraphics[width=0.98\columnwidth]{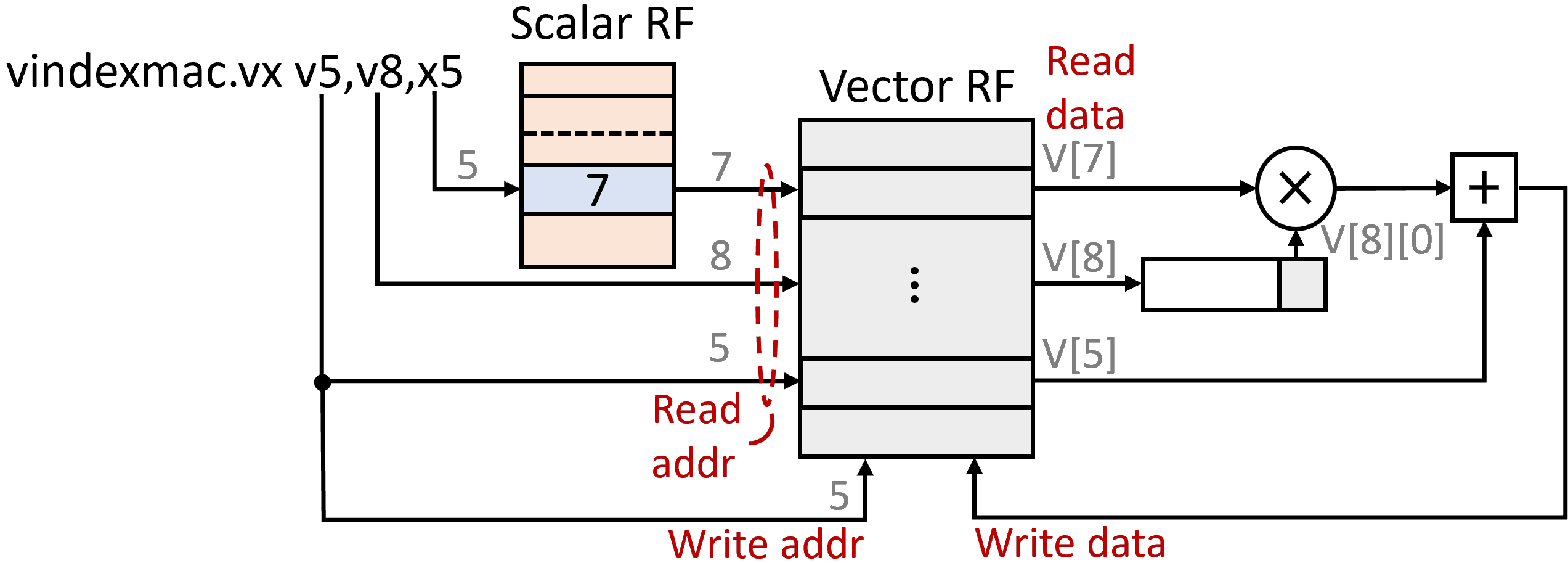}
\caption{The operation of the proposed {\tt vindexmac} instruction. The contents of the scalar register are used to address a specific vector register. The vector read is multiplied with the least significant element of another vector register that is read in parallel. 
The result of the multiplication is accumulated with the previous contents of the vector destination register.}
\label{f:indexmac}
\end{figure}

The term {\tt vrf} refers to the vector register file. The instruction has three operands: one vector destination register ({\tt vd}) and two source registers. One source register is a vector ({\tt vs2}) and the other is a scalar ({\tt rs}), as per the RISC-V ISA requirement for {\tt .vx} instructions. To execute {\tt vindexmac}, the scalar register rs is read and its contents (only the 5 least significant bits are actually needed) are used as an address to the vector register file. The content (i.e., value) of the vector register that is read via the address contained in {\tt rs} is multiplied with the least significant element of vector register {\tt vs2}. The operation of {\tt vindexmac} is visualized in Fig.~\ref{f:indexmac}.

\subsection{Sparse-Dense Matrix Multiplication using {\tt\bf vindexmac}}
\label{ss:multiplication_with_vindexmac}

Algorithm~\ref{a:vectorized_sparse-dense_indexmac} demonstrates the use of the new vindexmac instruction for a tile of $B$, with the non-zero values of a single row of matrix $A$ stored in a vector register. The essential differences between Algorithm~\ref{a:vectorized_sparse-dense_indexmac} and the previous algorithms are highlighted. As shown in lines 2 and 4, a tile of matrix $B$ is pre-loaded into the VRF. In line 6, the non-zero values of a row of matrix $A$ are loaded into the VRF, while the corresponding column indexes remain in the scalar RF. The use of the new {\tt vindexmac} instruction is shown in line 12. The end result is the elimination of the continuous vector loads of rows of matrix $B$, which take place in line 9 of Algorithm 3-S. 
Since {\tt vindexmac} uses only element 0 of vector register {\tt vs2}, a vector slide by one element position in {\tt vs2} is enough to proceed to the next non-zero element of the current row of matrix $A$ (Line 13 of Algorithm~\ref{a:vectorized_sparse-dense_indexmac}).

\begin{algorithm}[thb]
\caption{{Vector sparse-dense matrix multiplication with the use of the new {\tt vindexmac} instruction for \textbf{a tile} of $B$}}
    \label{a:vectorized_sparse-dense_indexmac}
	\begin{algorithmic}[1]
        \State set vector length
        \For{k=0 \textbf{until} k=L-1}
          \State {\color{blue1} vload vk, $B[k, :]$} \algorithmiccomment{preload $L$ rows of $B$}
        \EndFor
        \For{i=0 \textbf{until} i=num\_of\_rows\_of\_A-1}
            \State \textcolor{blue1}{vload vs2, values[i, :]}   \algorithmiccomment{load non zeros of $i$th row}
            \State \textcolor{blue1}{vload vd, $C[i, :]$} \algorithmiccomment{load the $i$th row of $C$}
            \For{j=0 \textbf{until} j=L*(N/M)-1}
                \State {block\_id =$\lfloor j/N \rfloor$}  \algorithmiccomment{calculate the block id}
                \State {load s1, col\_idx[i, j]}  \algorithmiccomment{load col. idx of $A$ to s1}
                
                \State {s1 += block\_id *M}\algorithmiccomment{calculate actual col idx}
                \State \textcolor{blue1}{vindexmac.vx vd, vs2, s1}
                \State \textcolor{blue1}{vs2 = vs2 $\gg 1$ }
                \algorithmiccomment{vector 1 slide to the right}
            \EndFor
            \State vstore vd, $C[i, :]$
            \algorithmiccomment{store to mem row i of $C$}
		\EndFor
  \end{algorithmic} 
\end{algorithm}

The number of rows $L$ of the tile of matrix $B$ that are loaded into the VRF and indexed by the {\tt vindexmac} instruction is limited by the number of available vector registers. In practice, we consider that a tile of matrix $B$ may consume half of the available vector registers, i.e., $L=16$. Due to the symmetry of matrix multiplication, the $L$ rows of matrix $B$ correspond to a set of $L$ columns of matrix $A$. Since matrix $A$ follows $N$:$M$ structured sparsity, it means that, in those $L$ columns, one can find only $L\cdot \left ( \frac{N}{M} \right )$ non-zero elements. Effectively, this property results in many more iterations in the outer loop (lines 5--16) of Algorithm~\ref{a:vectorized_sparse-dense_indexmac}, since the inner loop (lines 8--14) is executed for a small number of non-zero elements.

To improve utilization and to take advantage of the fact that the vector load for a row of matrix $A$ (line 6 of Algorithm~\ref{a:vectorized_sparse-dense_indexmac}) can bring -- at once -- $VectorLength$ elements in vector register {\tt vs1}, we need to consider more tiles of matrix $B$. This approach leads to a reformulation of Algorithm~\ref{a:vectorized_sparse-dense_indexmac}, which is shown in Algorithm~\ref{a:vectorized_sparse-dense_indexmac_tiling}. An intermediate loop is inserted (using index $m$), which is responsible to fetch and operate on the tiles of matrix $B$ that correspond to a complete row of non-zero elements of matrix $A$. The non-zero elements of a row of matrix $A$ loaded in line 3 of Algorithm~\ref{a:vectorized_sparse-dense_indexmac_tiling} are all processed by reloading the corresponding rows of matrix $B$ in phases. Specifically, the appropriate tiles of matrix $B$ are loaded in lines 6--8, in each phase. 

\begin{algorithm}[thb]
\caption{Vector sparse-dense matrix multiplication with the use of the new {\tt vindexmac} instruction by utilizing a complete row of $VectorLength$ non-zero elements of matrix $A$, which correspond to more than one tile of matrix $B$}
    \label{a:vectorized_sparse-dense_indexmac_tiling}
	\begin{algorithmic}[1]
        \State set vector length
        \For{i=0 \textbf{until} i=num\_of\_rows\_of\_A-1}
            \State vload vs1, values[i, :]  \algorithmiccomment{load values of row i}
            \State vload vd, $C[i, :]$
            \algorithmiccomment{load the $i$th row of $C$}
            \For{\textcolor{blue1}{m=0 \textbf{until} m=(M/N)*VL/L}}
                \For{\textcolor{blue1}{k=L*m \textbf{until} k=(m+1)*L-1}}
                  \State vload vk, $B[k, :]$ \algorithmiccomment{load a tile of $B$}
                \EndFor
                \For{\textcolor{blue1}{j=L*(N/M)*m \textbf{until} j=L*(N/M)*(m+1)-1}}
                    \State block\_id =$\lfloor j/N \rfloor$ \% L \algorithmiccomment{calculate the block id}
                    \State {load s1, col\_idx[i, j]}  \algorithmiccomment{load col. idx of $A$ to s1}
                    \State s1 += block\_id * M  \algorithmiccomment{calculate actual col idx}
                    \State vindexmac.vx vd, vs1, s1
                    \State vs1 = vs1 $\gg 1$ 
                    \algorithmiccomment{vector slide to the right}
                \EndFor
            \EndFor    
            \State vstore vd, $C[i, :]$
            \algorithmiccomment{store to mem row i of $C$}
		\EndFor
  \end{algorithmic} 
\end{algorithm}

The number of required reloading phases $m$ is determined by (a) the vector registers $L$ available for the loading of a tile of matrix $B$, (b) the sparsity pattern $N$:$M$, and (c) the maximum vector length of the processor $VL$. As stated earlier, in every $L$ columns of matrix $A$, we can find $L\, \left ( \frac{N}{M} \right )$ non-zero elements. Therefore, we need to take $m$ groups of $L$ columns of matrix $A$ to reach the $VectorLength$ $(VL)$ non-zero elements that can be present in one vector register, i.e.,  $m\, L\, \left ( \frac{N}{M} \right ) = VL$. Solving for $m$, we get that
\begin{equation}
m=\left ( \frac{M}{N} \right ) \left ( \frac{VL}{L} \right )
\end{equation}
This limit is used for the bounds of $m$ in line 5 of Algorithm~\ref{a:vectorized_sparse-dense_indexmac_tiling}. For the case of $1$:$4$ structured sparsity, and assuming that we allocate half of the vector register file for the tiles of matrix $B$ (i.e., $L$=16), then $m = (4/1)(VL/16) = VL/4$. It is crucial to point out that when the maximum value of $m$ is 1, Algorithm~\ref{a:vectorized_sparse-dense_indexmac} is preferred over Algorithm~\ref{a:vectorized_sparse-dense_indexmac_tiling}. When $m=1$, there is no need to interchange the tile of matrix $B$ that is loaded in the vector register file. Instead, Algorithm ~\ref{a:vectorized_sparse-dense_indexmac} would, by construction, avoid the continuous loads that refer to the same rows of matrix $B$. 

To further improve performance, the sparse$\times$design matrix multiplication kernel that uses the newly proposed {\tt vindexmac} instruction (as shown in Algorithm~\ref{a:vectorized_sparse-dense_indexmac_tiling}) could be unrolled across three dimensions: (a) the elements of the same row of matrix $A$ (inner loop); (b) the tiles of matrix $B$ (intermediate loop); and (c) across the various output rows (outer loop). 
The proposed approach, which preloads a tile of matrix $B$ that utilizes a significant portion of the VRF,  
limits the depth of loop unrolling. To optimally apply unrolled sparse-dense matrix multiplication with the {\tt vindexmac} instruction, a balanced approach is required, combining unrolling and appropriate tiling of $B$ to efficiently utilize the 32 vector registers of a RISC-V architecture.

\subsection{Tiled execution for larger matrices $A$ and $B$}
\label{s:tiling-vindexmac}
Algorithm 6 can work on a matrix $A$ with any number of rows. However, it assumes that the non-zero elements of $A$ loaded into a vector register at each step correspond to at most $m\times L$ columns. This is the reason for loading $m$ tiles of $B$ with $L$ rows each at each round. Additionally, Algorithm 6 assumes that all columns of $B$ can fit into one vector register.

\begin{figure}[t]
\includegraphics[width=0.98\columnwidth]{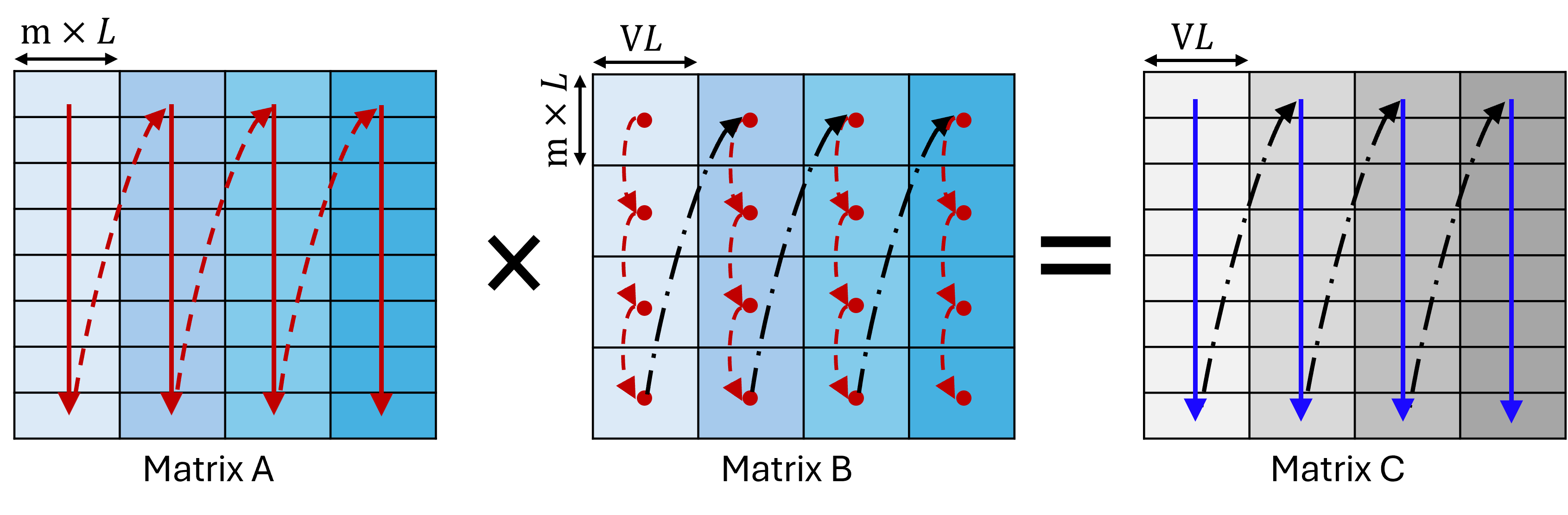}
\caption{The order of execution of Alg. 6 across vertical segments of $A$ and $B$.}
\label{f:dataflow}
\end{figure}

In the general case, where the columns of A (and consequently the rows of B) exceed 
$m\times L$, and the columns of $B$ are larger than $VL$, we need to apply Algorithm 6 repeatedly using the dataflow shown in Fig.~\ref{f:dataflow}. 
For one vertical segment of $VL$ columns of $B$, we process all rows of $A$ in groups of $m\times L$ columns. 
In particular, we first process all rows of $A$ that belong to the same vertical segment of $m\times L$ columns before moving to the next segment. Within each vertical segment of $A$, we follow the execution order of Algorithm 6, keeping one group of $m\times L$ rows of $B$ stationary. Once all vertical segments of A are processed, we repeat the process for subsequent vertical segments of B by fetching the next $VL$ columns and restarting.

This tiled dataflow can be employed also when executing sparse-dense matrix multiplication to a multicore environment. The computation for every vertical segment of $B$ consisting of $VL$ columns can be assigned to a different core. This partitioning, allows each core to operate 
on independent segments of the output matrix without facing any  conflicts. However, with this dataflow, the elements of structured-sparse matrix $A$ are accessed by all cores in parallel. The benefits of implementing sparse-dense matrix multiplication using {\tt vindexmac} in a multicore setup under this dataflow is quantified in Section~\ref{ss:multicore}.

\subsection{Hardware support for {\tt vindexmac} execution}
The vector register file in the RISC-V ISA is instrumental in supporting efficient vector processing. It consists of 32 vector registers, each capable of storing a large volume of data. Each register's length is an implementation-defined constant parameter decided by the processor designer. To support all available vector instructions, including three-operand instructions (e.g., {\tt vmacc}, {\tt vmadd}, etc.), the vector register file is required to have three read ports and one write port.

To execute the {\tt vindexmac} instruction, we need to access both the scalar and the vector register files. This is an inherent attribute of all RISC-V scalar-vector instructions that have the letter `{\tt x}' in the suffix, e.g., {\tt .vx}, {\tt .vxm}, {\tt .wx}, etc.

In this work, we target vector processors that follow a so called \textit{decoupled} architecture, which consists of a scalar core that is responsible for instruction fetching and orchestration of execution, and a vector engine that executes the vector operations received from the scalar core~\cite{tarantula, xuantie, cornell-vector, ara, riscv-vector, vitruvius}. Since, at any given time, there may be many vector instructions that require the values of scalar registers, the scalar core is responsible to transfer these values to the vector processor, together with the vector instructions themselves. 

Therefore, the value of the scalar register {\tt rs} required by {\tt vindexmac} to address the vector register file is already provided by the scalar core. 
The given address drives one of the read ports of the vector register file, which outputs the requested vector operand. Another read port is used to read the elements of {\tt vs2} and the third one reads {\tt vd}, as required by all multiply-and-accumulate scalar-vector instructions already present in the RISC-V ISA. Therefore, the only hardware requirement of the new {\tt vindexmac} instruction is the addition of a multiplexer in front of the address bus of one of the read ports of the vector register file, which selects between {\tt vs1} for normal vector arithmetic operations, or the 5 least significant bits of {\tt rs} (as required by {\tt vindexmac}). In other words, by re-using the hardware infrastructure of existing scalar-vector multiply-add instructions, the new instruction does \textit{not} require an additional read port in the vector register file. Instead, it merely requires a 5-bit 2-to-1 multiplexer in front of an existing read port.

Specifically, a 32-register VRF, with each register being 512 bits wide (VL=16 for 32-bit words), with 3 read and 1 write port, synthesized with Cadence Genus at the 28 nm technology node for a 1 GHz clock period, requires a hardware area of 56,527 $\mu m^2$. The additional area needed to accommodate the address multiplexing logic is just 4 $\mu m^2$ extra.

Similarly, in fully integrated scalar-vector architectures~\cite{m3, arm-sve}, 
the {\tt vindexmac} instruction would be implemented in exactly the same manner as any of the other {\tt .vx} instructions in a fully integrated scalar-vector setup. The only difference is that the scalar value provided would be used to drive one of the read ports of the vector register file, rather than directly participating in the computation.

Being a {\tt .vx} instruction, {\tt vindexmac} follows the standard encoding dictated by the RISC-V ISA for scalar-vector instructions~\cite{rvv}. 
However, it terms of the instruction functionality, there is a slight deviation from typical scalar-vector operations. The vector register identifier {\tt vs2} is used only for its least significant element {\tt vs2[0]}, i.e., it plays the role of the scalar value. The actual vector is fetched via an indirect read using the 5 least significant bits of {\tt rs} as an address into the vector register file.

\section{Experimental Evaluation}
\label{s:eval}

The experimental evaluation presented in this section has two goals. The first one is to identify the performance limits of the introduced row-based formulation of sparse-dense matrix multiplication using only the current definition of the RISC-V ISA; i.e., using only existing instructions. The evaluation includes the execution of three \textit{pruned} state-of-the-art CNNs -- Resnet50~\cite{resnet}, DenseNet121~\cite{densenet} and InceptionV3~\cite{inception} -- under various loop unrolling scenarios on a superscalar out-of-order processor augmented with a decoupled vector unit. All calculations in all three CNNs are performed in the fp32 data format. The pruned CNNs employ structured block sparsities of 1:4 and 2:4. Note that, as demonstrated in~\cite{learning-n-m} and~\cite{nvidia-block-sparse}, the selected structured-sparsity patterns do not change the classification accuracy of the specific CNNs, as compared to their fully dense counterparts.

After identifying the configuration that offers the greatest speedup when using the current RISC-V ISA, the obtained performance is compared with the performance achieved by the structured-sparse matrix multiplication kernels that employ the proposed new {\tt vindexmac} instruction. For the latter case, various alternatives are also explored and the one that achieves the lowest execution time is used in the comparisons.

\begin{table}[t]
\centering
\caption{Simulated Processor Configuration}
\label{t:proc-det}
\renewcommand\tabularxcolumn[1]{m{#1}} 
\begin{tabularx}{\columnwidth}{cX}
\hline
Scalar core & 
\begin{itemize}[leftmargin=6pt]
\item 
RISC-V ISA (RV64GC), 8-way-issue out-of-order,
16-entry LSQ, 90 physical integer and 90 physical
 floating-point registers, 60-entry ROB
\item
L1I cache: 1-cycle hit latency, 4-way, 64KB
\item 
L1D cache: 2-cycle hit latency, 4-way, 64KB
\item 
L1I and L1D have 64B cache line
\end{itemize}\\
\hline
Vector engine & \begin{itemize}[leftmargin=6pt]
\item 512-bit vector engine with 16-lane configuration (32-bit elements $\times$ 16 execution lanes)
\item The vector engine is connected directly to the L2 cache through 16 store queues and 16 load queues
\end{itemize} \\
\hline
L2 cache & \begin{itemize}[leftmargin=6pt]
\item 8-way, 8-bank
\item 8-cycle hit latency, 512KB shared by both the big core and the vector engine
\item 64B cache line
\end{itemize} \\
\hline
Main Memory & DDR4-2400, 19.2 GB/s memory bandwidth\\ 
\hline
\end{tabularx}
\end{table}

\begin{figure*}[thb]
\centering
\includegraphics[width=2\columnwidth]{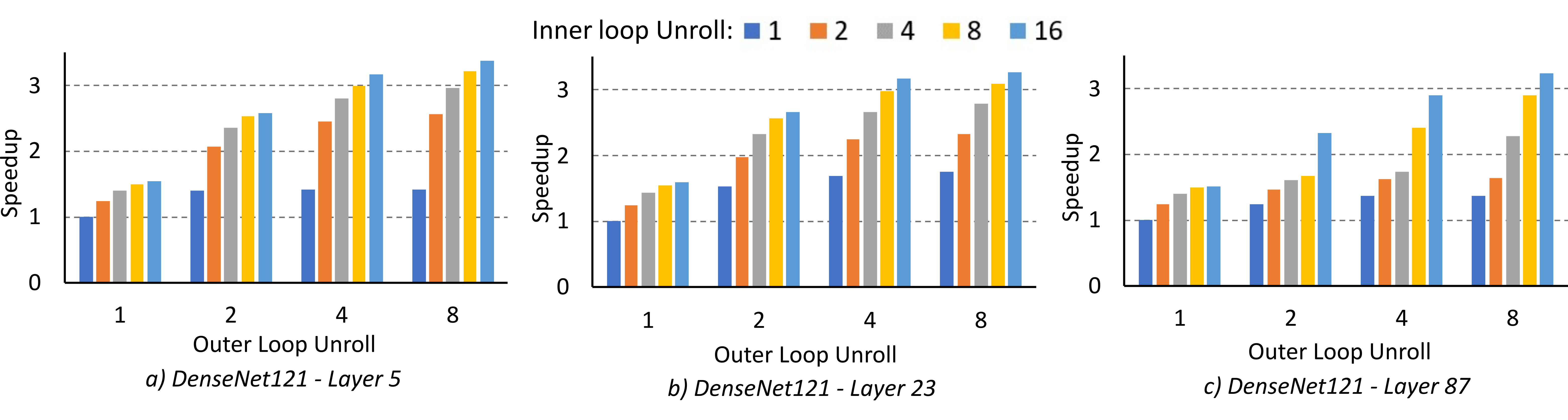}
\caption{The performance of the optimized sparse-dense matrix multiplication using the current definition of the RISC-V ISA (Algorithm 3-S) used for the execution of three CNN layers of Densenet121~\cite{densenet}, assuming a 1:4 sparsity pattern: (a) Layer 5; (b) Layer 23; and (c) Layer 87. The reported cases examine the combined effect of the unrolling of the inner and outer loops of Algorithm 3-S. The speedup is reported relative to the performance of the rolled variant.}
\label{f:Unrolling-RISCV-ISA-only}
\end{figure*}

\subsection{Experimental setup}
\label{ss:exp-setup}

For all experiments, we utilize a fully implemented decoupled vector unit connected to an out-of-order superscalar processor, i.e., model 1bDV in~\cite{cornell-vector}. 
The choice of a high-end superscalar core (connected to the vector engine) aims to safeguard that the non-vector side of the processor does not skew the performance results of the vector engine. In other words, the superscalar processor helps to ``isolate'' the performance of the vector unit by ensuring that there are no ``bottlenecks'' on the non-vector side. As such, the reported results highlight purely the efficacy of the new vector framework, without any indirect effects/noise from the scalar side.

This design was modeled in the gem5 simulator~\cite{gem5-orig, gem5-2020} and the salient design parameters of the simulated processor setup are summarized in Table~\ref{t:proc-det}. The new {\tt vindexmac} instruction was incorporated in the RISC-V GNU toolchain and its operation was implemented in the decoupled vector engine in gem5.

The convolutions of each layer of the examined CNNs are mapped to sparse-dense matrix multiplications $A\times B$~\cite{nvidia-block-sparse}. Matrix $A$ includes the structured-sparse weights and matrix $B$ the input features of the corresponding CNN layers. Input features are considered dense, since there is no clear statistical attribute that can be exploited for them. Even if part of the input features contain zero values generated by the corresponding ReLU activation functions in each layer, their number is highly sensitive to the actual input values and filter weights.

\subsection{Identifying peak performance using only instructions from the current RISC-V ISA}
\label{ss:current-ISA}

Our initial goal is to examine which of the approaches highlighted in Algorithms 1S--3S (that use only the current definition of the RISC-V ISA) offer the best scaling in execution time for various Sparse-dense matrix multiplications and and compare also the best of them to other state-of-the-art approaches.

As discussed in Section~\ref{ss:loop-unrolling}, the inner and the outer loops of all algorithms can be unrolled to increase performance. Outer-loop unrolling corresponds to processing multiple rows of sparse matrix $A$ in parallel, while inner-loop unrolling enables the simultaneous processing of multiple non-zero elements of the same row of matrix $A$, in a flattened form. To highlight how unrolling can improve runtime, we perform several experiments using three representative layers from the DenseNet121 CNN~\cite{densenet}, assuming 1:4 sparsity. 

The effect of optimized unrolling presented in Section~\ref{ss:loop-unrolling} is examined first for Algorithm 3-S.
The speedup achieved in each case, relative to the rolled version of Algorithm 3-S, is shown in Fig.~\ref{f:Unrolling-RISCV-ISA-only}. Each bar corresponds to a different inner-loop unroll factor, while outer-loop unrolling factors are represented by the different sets of bars along the x-axis. For instance, the unrolled example shown in Algorithm~\ref{a:vectorized_row-wise_sparse-dense-scalar-unrolled-vrgather} of Section~\ref{ss:loop-unrolling} corresponds to the case of using an unrolling factor of 2 for both the inner and outer loops. As demonstrated in Fig.~\ref{f:Unrolling-RISCV-ISA-only}, increasing unrolling improves performance in all cases, reaching a speedup of more than $3\times$ relative to the baseline implementation. 

Speedup is maximized when the inner loop can process 16 non-zero elements in parallel and the outer loop is unrolled by a factor of 8. the factors are constrained by the ``size'' of the the vector unit. The inner loop unroll factor of 16 corresponds to the hardware $VectorLength$ shown in Table~\ref{t:proc-det}. These specific unrolling factors are directly linked to the hardware; i.e., 
the factors are constrained by the ``size'' of the the vector unit. The inner loop unroll factor of 16 corresponds to the hardware $VectorLength$ shown in Table~\ref{t:proc-det}. Similarly, the unrolling factor of 8 of the outer loop is limited by the total number of vector registers.

\begin{figure}
    \centering
    \includegraphics[width=0.85\linewidth]{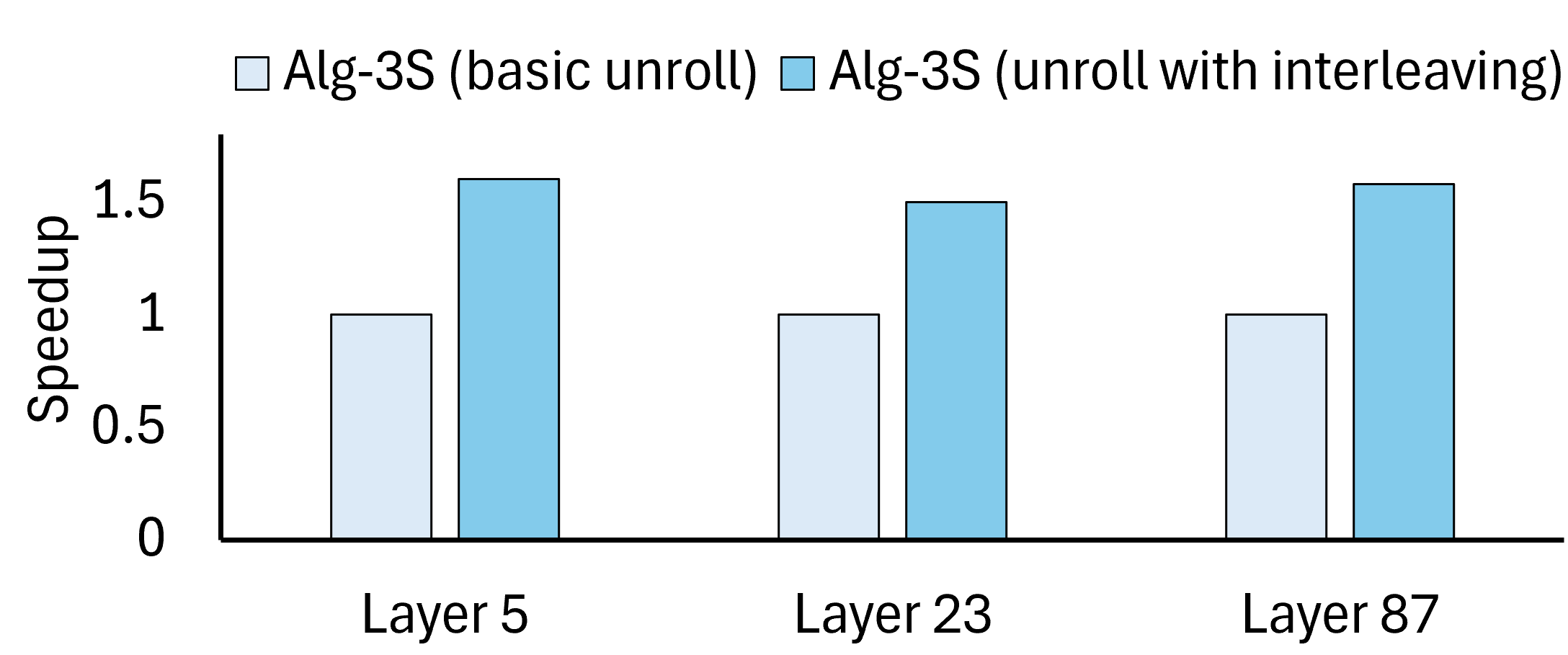}
    \caption{Speedup of interleaved unrolling on Alg-3S, normalized to the speedup of Clang's~\cite{clang} basic unrolling methodology with inner and outer loop unrolling factors of 16 and 8, respectively.}
    \label{f:unroll-compare}
\end{figure}

This significant gain is the combined result of (a) the structure of the row-based matrix multiplication algorithm, (b) the hybrid data placement across the vector and the scalar register files, and (c) the efficiency of interleaved loop unrolling shown in Section~\ref{ss:loop-unrolling} that reduces unnecessary name dependencies and allows sharing of index calculations.
The benefit of interleaving the execution of the operations of the unrolled inner loop has been quantified separately and compared to the traditional unrolling that can be performed automatically by Clang~\cite{clang} The results for the same CNN layers and the same inner and outer loop unrolling factors are shown in Fig.~\ref{f:unroll-compare}. In all cases, the interleaving presented in this work achieves 50\% higher performance.

\begin{figure}
\centering
\includegraphics[width=0.85\columnwidth]{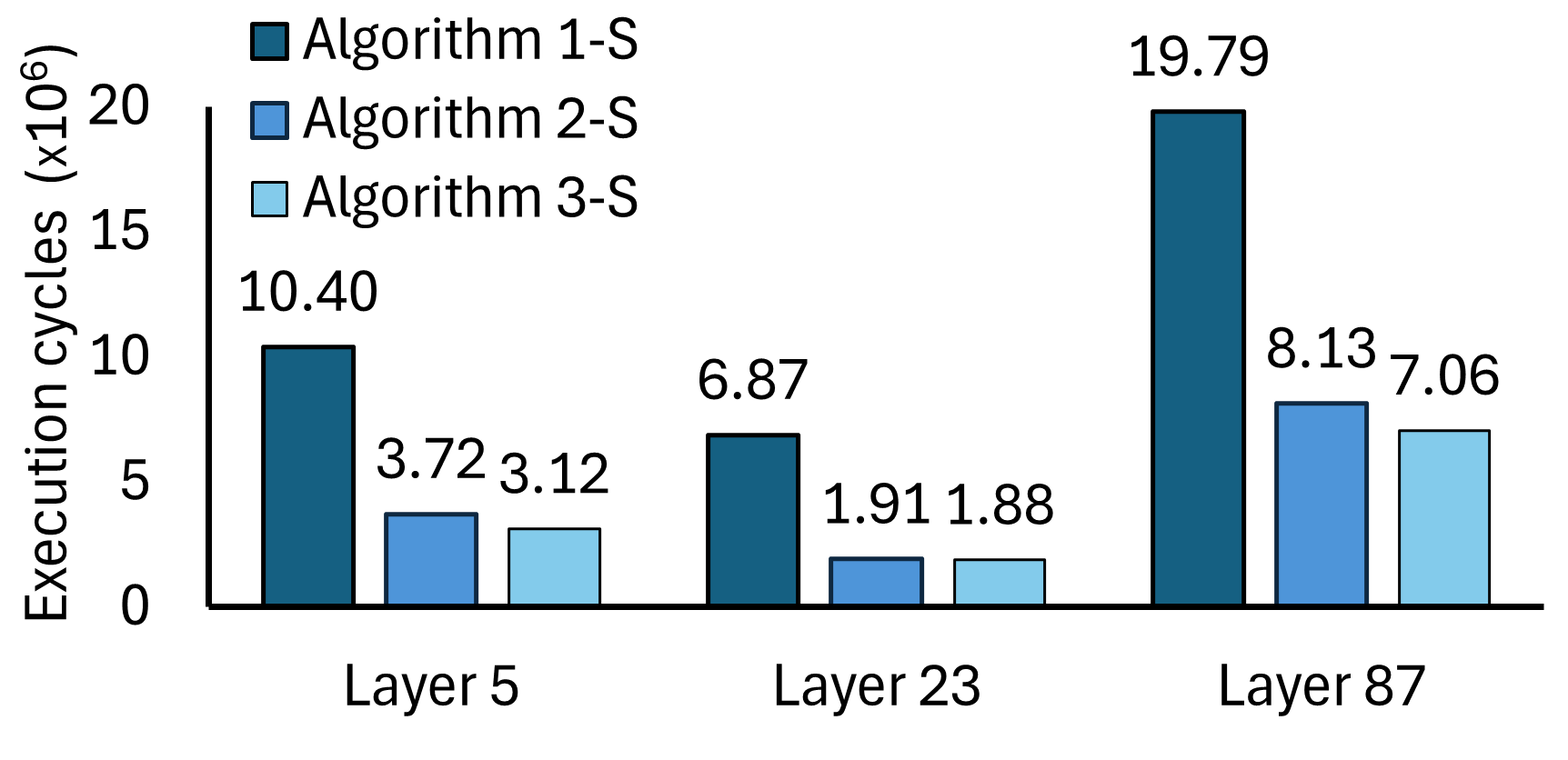}
\caption{{Comparison of the unrolled versions of Algorithms 1-S, 2-S, and 3-S, which achieve the best performance in each case for the same outer and inner loop unrolling configuration.}}
\label{f:compare_alg}
\end{figure}

Algorithms 1-S, 2-S, and 3-S can all equally benefit from unrolling with interleaving. Fig.~\ref{f:compare_alg} illustrates the execution time of each algorithm for the same CNN layers when the inner loop processes 16 non-zero elements in parallel and the outer loop is unrolled by a factor of 8. Algorithm 1-S exhibits the lowest performance, due to the overhead associated with transferring data between vector and scalar registers for each operation. In contrast, Algorithms 2-S and 3-S, which employ a mixed index and data placement across scalar and vector registers, achieve better performance. Algorithm 2-S uses scalar registers to store both the non-zero values and column indices of matrix $A$, increasing the demand on scalar registers to maintain this data locally. Conversely, Algorithm 3-S stores the non-zero elements of $A$ in vector registers. This approach eliminates the need for independent scalar load instructions for each non-zero element of $A$, as required by Algorithm 2-S. This change is the reason for the slight superiority of Alg 3-S relative to Alg. 2-S.

\begin{figure}
\centering
\includegraphics[width=0.85\columnwidth]{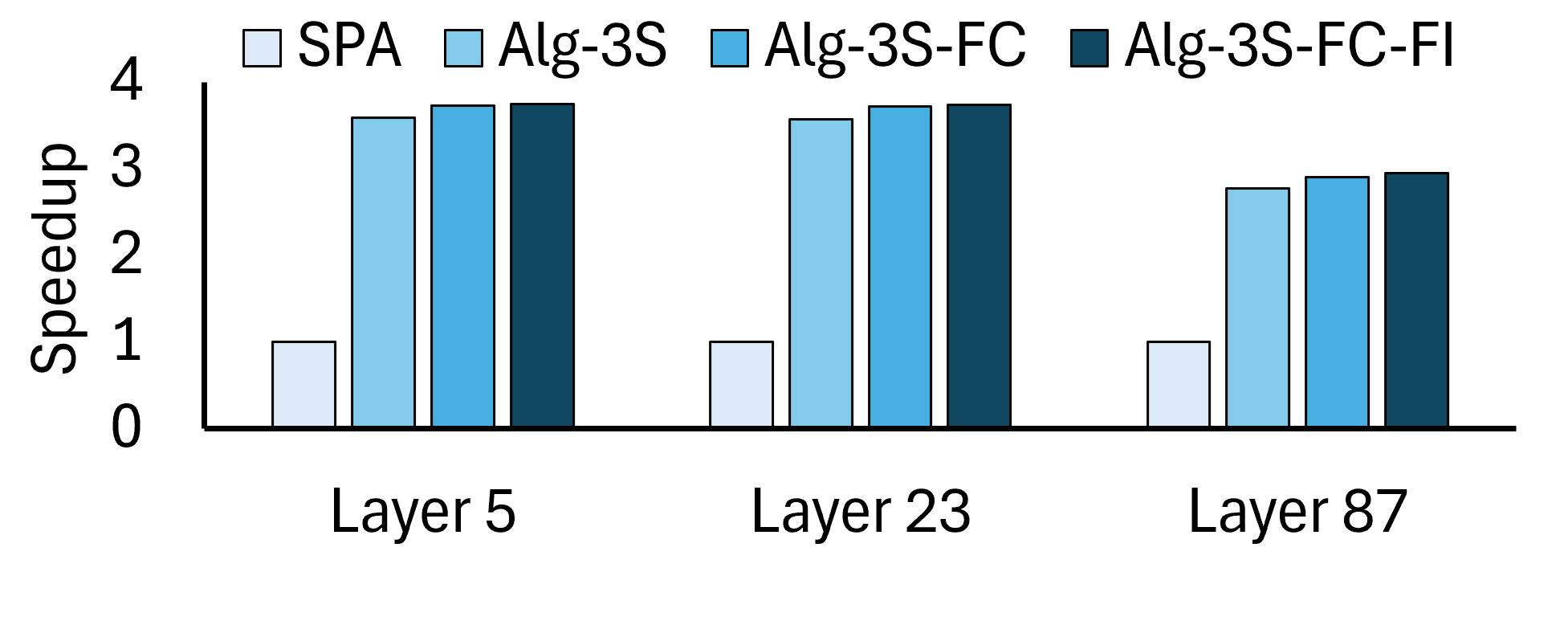}
\caption{Performance analysis of the SPA, Alg-3S, Alg-3S-FC, and Alg-3S-FC-FI implementations, each normalized relative to the SPA implementation.}
\label{f:spa_vs_rest}
\end{figure}

To identify a state-of-the-art Sparse-dense baseline to compare with the implementation of {\tt vindexmac}, we first compare the best configuration identified so far for Alg-3S, with various other alternatives:
\begin{itemize}
\item {Alg-3S-FC (Full Column): It enhances Alg-3S by leveraging full column indexes to avoid the additional overhead in generating the columns themselves, as required by the structured sparse blocks. This approach is effectively equivalent to a CSR implementation, with the \emph{added simplification} that the number of elements per row is predetermined, as in structured sparsity. These broader column indexes are anticipated to improve runtime, albeit with an increase in storage requirements.}
\item Alg-3S-FC-FI (Full column with Fast Indexing): 
The optimized Alg-3S-FC can be further enhanced by incorporating the custom instructions proposed in~\cite{islped}, which integrate/fuse the address generation phase into both scalar and vector load instructions. This acceleration of indexing calculations in hardware is orthogonal to all approaches presented in this work and can be equally applied to {\tt vindexmac} extensions as well. 
\item {SPA:
SPA is the most widespread algorithm that handles sparse matrix multiplication when both arrays $A$ and $B$ are substantially sparse~\cite{spa,generic-spgemm, upc-spgemm}. SPA is included for completeness even if its main target is sparse-sparse matrix multiplications with high degrees of sparsity.}
\end{itemize}
{As shown in Fig.~\ref{f:spa_vs_rest}, both Alg-3S-FC and Alg-3S-FC-FI yield performance improvements, though not substantial over Alg-3S. This is primarily because the indexing arithmetic required by Alg-3S -- necessitated by the compact column indexes of the structured sparse format -- can be shared. Combined with reduced name dependencies under loop unrolling, this has only a minor effect on performance when executed on an Out-of-Order (OoO) superscalar core.  More importantly, however, the utilization of full columns (as in Alg-3S-FC and Alg-3S-FC-FI) results in a significant storage overhead of 14.7\%, 26.5\%, and 23.5\% for the sparse matrix in Layers 5, 23, and 87, respectively, which is an undesired side effect. The limited performance of the SPA implementation is attributed to its use of vector-indexed memory operations to manage non-zero elements. These operations adversely affect the overall performance; they lead to a high number of cache misses, even with a relatively small number of vector instructions.}

Given (a) the target of structured sparsity, (b) the smaller column indexes, and (c) the requirement to maintain full RISC-V ISA compliance, we select Alg-3S as the reference baseline, henceforth called `SpMM' in the experiments. More specifically, we will shortly describe the best setup as SpMM(16,8), referring to an 
inner-loop unrolling of 16 and an outer-loop unrolling of 8.

It is important to note that if we were to select Alg-3S-FC-FI as the baseline, which employs full columns and the custom instructions proposed by~\cite{islped}, the performance speedups with the proposed methodology would only be marginally smaller (as indicated by the marginal performance improvement of Alg-3S-FC-FI over Alg-3S in Fig.~\ref{f:spa_vs_rest}). The speedups achieved by the proposed approach when compared to Alg-3S-FC-FI will be quantified experimentally in the following sub-section.

 \begin{figure*}
\centering
\includegraphics[width=1.85\columnwidth]{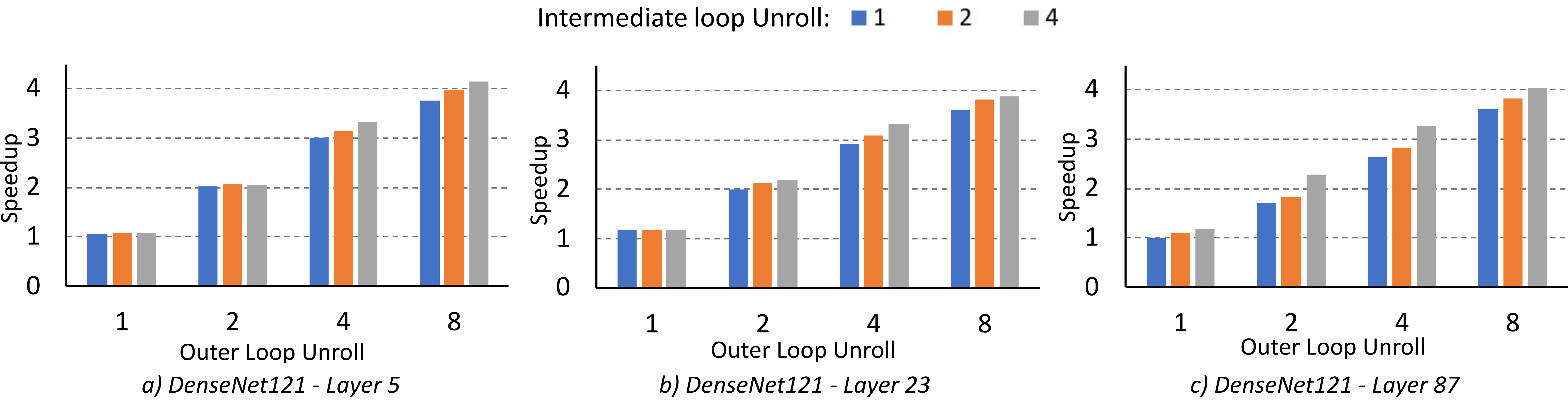}

\caption{The performance of the optimized sparse-dense matrix multiplication, enhanced with the proposed {\tt vindexmac} instruction (Algorithm~\ref{a:vectorized_sparse-dense_indexmac_tiling}). Results are provided for the execution of three CNN layers of Densenet121~\cite{densenet}, assuming a 1:4 sparsity pattern: (a) Layer 5; (b) Layer 23; and (c) Layer 87. The reported cases examine the combined effect of the unrolling of the intermediate and outer loops of Algorithm~\ref{a:vectorized_sparse-dense_indexmac_tiling}. The speedup is reported relative to the performance of the rolled variant.}
\label{f:Unrolling-vindexmac}
\end{figure*}

\subsection{Identifying the best configuration when using the new {\tt vindexmac} instruction}
\label{ss:best-vindexmac}

Sparse-dense matrix multiplication using the new {\tt vindexmac} instruction has been evaluated for various configurations that execute the same three layers of DenseNet121~\cite{densenet} with 1:4 structured sparsity. 
The proposed kernel evaluated is the one shown in Algorithm~\ref{a:vectorized_sparse-dense_indexmac_tiling}, which extends the row-based matrix multiplication with a mixed placement of data in the vector and scalar register files and the new instruction  
that operates on local data and reduces memory traffic.
Algorithm~\ref{a:vectorized_sparse-dense_indexmac_tiling} consists of three distinct loops (inner, intermediate, and outer) that can be arbitrary unrolled. To identify the best configuration, we choose to always fully unroll the inner loop, and then investigate the effect of separately unrolling the other two loops. In all examined cases, the tile of matrix $B$ is composed of $L$=16 rows that occupy half of the vector register file. The value of $L$ and the sparsity pattern $N$:$M$ determine the number of iterations of the intermediate loop and the effective unrolling of the inner loop.

The achieved speedup for each examined configuration is shown in Fig.~\ref{f:Unrolling-vindexmac}. To have a consistent view of the achieved speedup, the results of Fig.~\ref{f:Unrolling-RISCV-ISA-only} and Fig.~\ref{f:Unrolling-vindexmac} have been normalized to the same basis, i.e., the runtime of the rolled version of Algorithm 3-S, which uses instructions from the current definition of the RISC-V ISA. From the speedups reported in Fig.~\ref{f:Unrolling-vindexmac}, we can observe that the unrolling of the outer and intermediate loops substantially improves the performance. In all cases, the speedup is maximized when the outer loop unrolling factor is 8 and the intermediate loop has been unrolled by a factor of 4. The outer-loop unrolling corresponds to the number of rows of sparse matrix $A$ that are computed in parallel. Since a tile of matrix $B$ occupies half of the vector register file, only the other half of the vector register file is available to facilitate higher degrees of loop unrolling. Thus, an unroll factor of 8 is the maximum possible, based on the number of available vector registers. An intermediate loop-unrolling factor of 4 corresponds to a full unroll for the examined configuration, i.e., $VectorLength$=16 elements (see Table~\ref{t:proc-det}), $L$=16 and $N$:$M$=1:4. 

In the rest of the experimental results, we use the best configuration for the proposed approach, as identified in Fig.~\ref{f:Unrolling-vindexmac}: intermediate-loop unrolling of 4 and outer-loop unrolling of 8. We define this setup as Proposed(8,4).

\begin{figure}
\centering
\includegraphics[width=0.72\columnwidth]{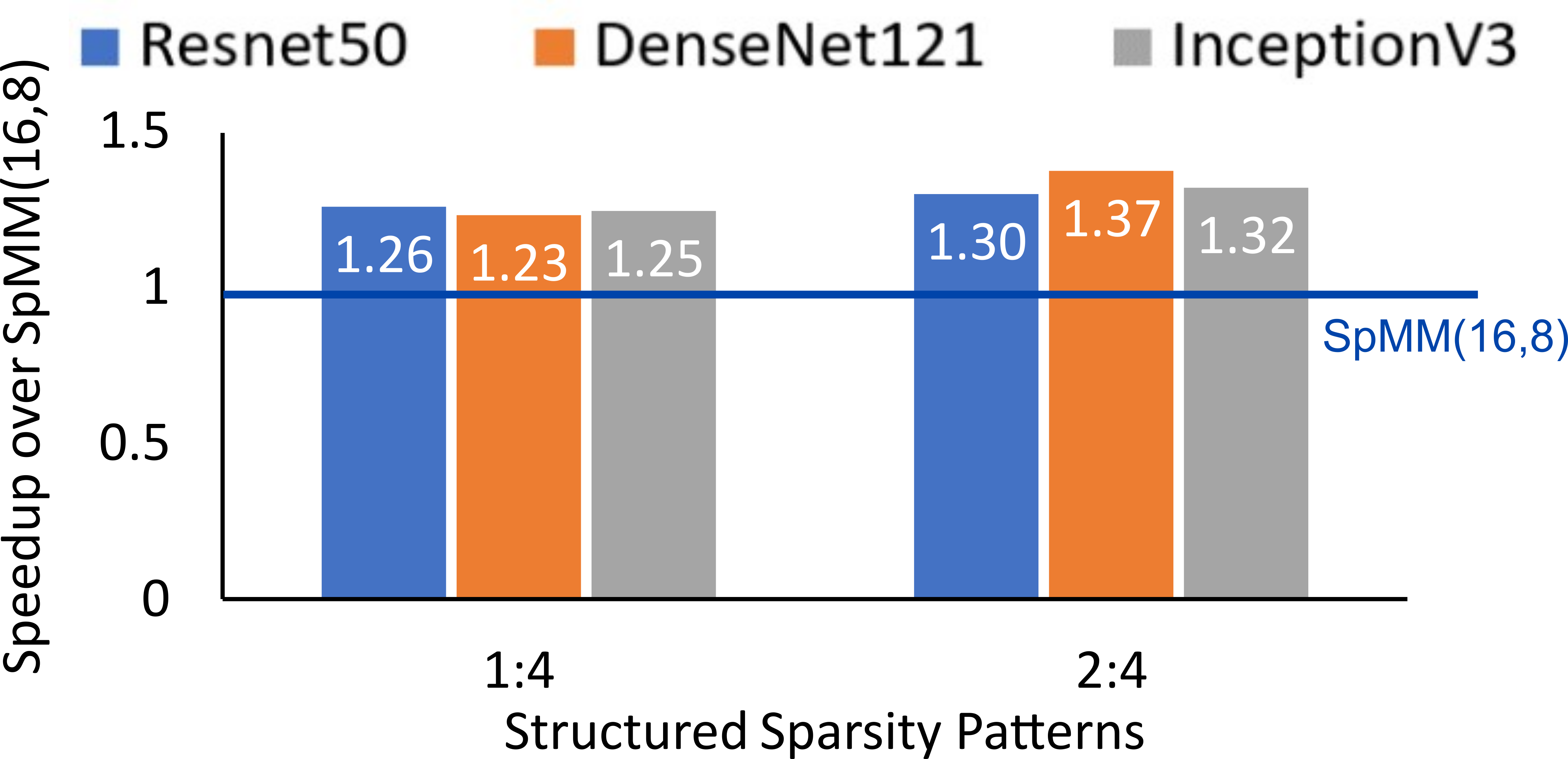}
\caption{The speedup achieved by the proposed approach (`Proposed(8,4)') in the \textit{total} execution times of the three examined CNNs. Results are shown for (a) 1:4, and (b) 2:4 structured sparsity, normalized to `SpMM(16,8)' of the respective sparsity.}
\label{f:total-speedup}
\end{figure}

\subsection{Comparisons}

The best identified configurations derived for SpMM (in Section~\ref{ss:current-ISA}) and for the proposed approach (in Section~\ref{ss:best-vindexmac}) are hereby compared when executing the three complete CNN models. Fig.~\ref{f:total-speedup} illustrates the achieved speedup of the proposed approach (`Proposed(8,4)') over `SpMM(16,8)' in the three evaluated CNNs, assuming 1:4 and 2:4 sparsity patterns. Clearly, the performance of `Proposed(8,4)' is substantially better in all cases. Across all three CNNs, the average speedup achieved for 1:4 sparsity is 1.25$\times$, which increases to 1.33$\times$ for 2:4 sparsity. As expected, the average speedups of the proposed approach when compared against Alg-3S-FC-FI -- with the same unroll factors for both the inner and outer loops -- are only marginally smaller: 1.20$\times$ for 1:4 sparsity and 1.28$\times$ for 2:4 sparsity.

It should be stressed that this extra 25\%--33\% improvement offered by the proposed new {\tt vindexmac} instruction is \textit{extra} speedup, \textit{over and above} the already highly optimized row-based kernel introduced in this paper. The \textit{overall} improvement corresponds to above $4\times$ speedup, as compared to baseline approaches performing sparse-dense matrix multiplication. This result highlights the combined improvement offered by mixed-data placement in both the scalar and vector register files and the reduced memory traffic achieved by the new instruction. 

\begin{figure}
\centering
\includegraphics[width=0.72\columnwidth]{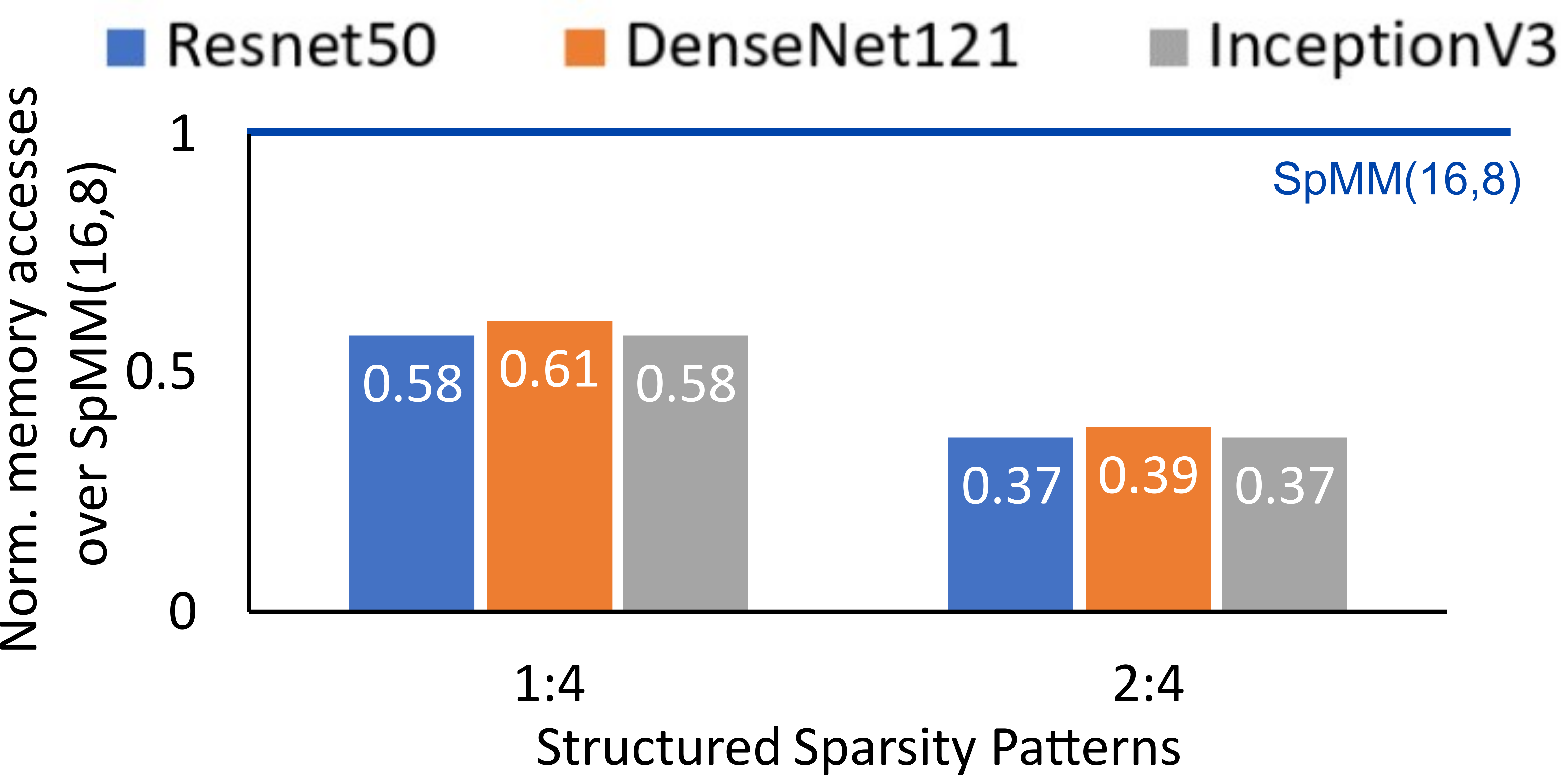}
\caption{The normalized number of \textit{total} memory accesses observed when using the proposed approach ('Proposed(8,4)') in the three examined CNNs, for (a) 1:4, and (b) 2:4 structured sparsity. The normalization is with respect to `SpMM(16,8)' of the respective sparsity.}
\label{f:Total-mem-accesses}
\end{figure}

In general, the above-mentioned improvements in performance when using the `Proposed(8,4)' approach are reaped from (a) the reduction of vector operations per iteration, and (b) the elimination of unnecessary vector loads from memory and their transformation into indirect reads from the vector register file (as facilitated by the {\tt vindexmac} instruction). By pre-loading tiles of matrix $B$ into the vector register file, the proposed approach exploits data locality very effectively, thereby lowering the memory traffic. The results presented in Fig.~\ref{f:Total-mem-accesses} quantifies this reduction in total memory accesses when using the proposed {\tt vindexmac} instruction. The presented results are normalized to the number of memory accesses observed with `SpMM(16,8)' of the respective sparsity. As can be seen, the total memory accesses decrease markedly. For sparsity 1:4, the memory accesses are reduced by 42\%, on average, while the average reduction for 2:4 sparsity is 63\%.

\subsection{Performance scaling: Increasing hardware vector length or employing multiple cores}
\label{ss:multicore}

To further explore the scalability of the proposed approach built around the new {\tt vindexmac} instruction, we experiment with different hardware configurations. Specifically, we investigate the effect on performance when scaling the available parallelism of the vector unit. Recall that the basic configuration (as presented in Table~\ref{t:proc-det}) utilizes 512b vector registers to operate across 16 parallel lanes, assuming 32-bit words. 
Hence, in the basic configuration, each vector register has 16 vector elements, i.e., $VectorLength$=16. 
In addition to the basic hardware configuration, we hereby examine two more configurations: 256b vector registers with 8-lane parallelism ($VectorLength$=8), and 1024b vector registers with 32-lane parallelism ($VectorLength$=32).

\begin{figure}
\centering
\includegraphics[width=0.95\columnwidth]{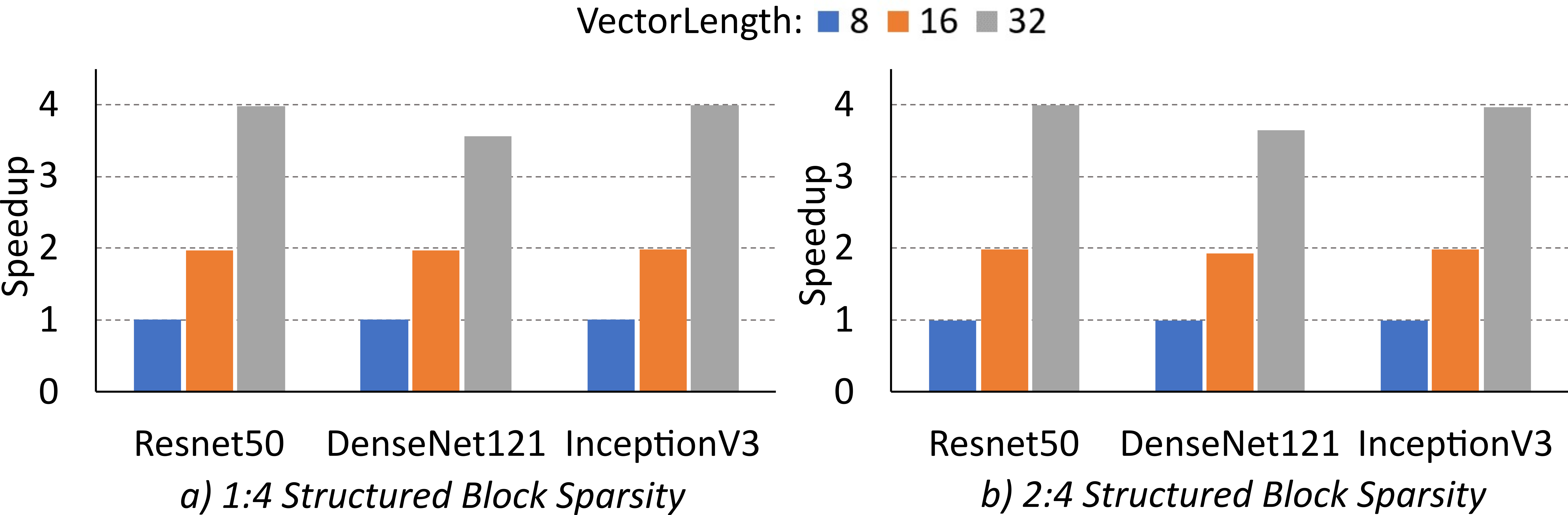}
\caption{The effect of hardware vector length on the proposed approach in the three examined CNNs, for (a) 1:4 and (b) 2:4 structured sparsity. 
The normalization of the results is with respect to the $VectorLength$=8 configuration of the respective sparsity.}
\label{f:different-vl}
\end{figure}

The execution time measured for the three CNNs -- assuming the same two sparsity patterns -- is shown in Fig.~\ref{f:different-vl}. The execution time in each case is normalized to the execution time of the $VectorLength$=8 configuration of the respective sparsity.  
As expected, when the vector length increases, the reaped speedup increases proportionally. Since the longer vector length is accompanied by more execution lanes, it means that more columns of the output matrix $C$ are being processed in parallel at any given time. This, in turn, leads to significantly lower execution times. 
As can be seen in Fig.~\ref{f:different-vl}, in most cases, the achieved speedup over $VectorLength$=8 is near-perfect, i.e., speedups approaching 2$\times$ with $VectorLength$=16 and speedups approaching 4$\times$ with $VectorLength$=32.
This desired behavior is aided by the L2 cache, which is big enough in the current processor configuration to accommodate the bigger volume of data as the vector length increases. Naturally, the performance improvement would saturate if the working data size were to approach the L2 cache capacity.

\begin{figure}[t]
\centering
\includegraphics[width=0.65\columnwidth]{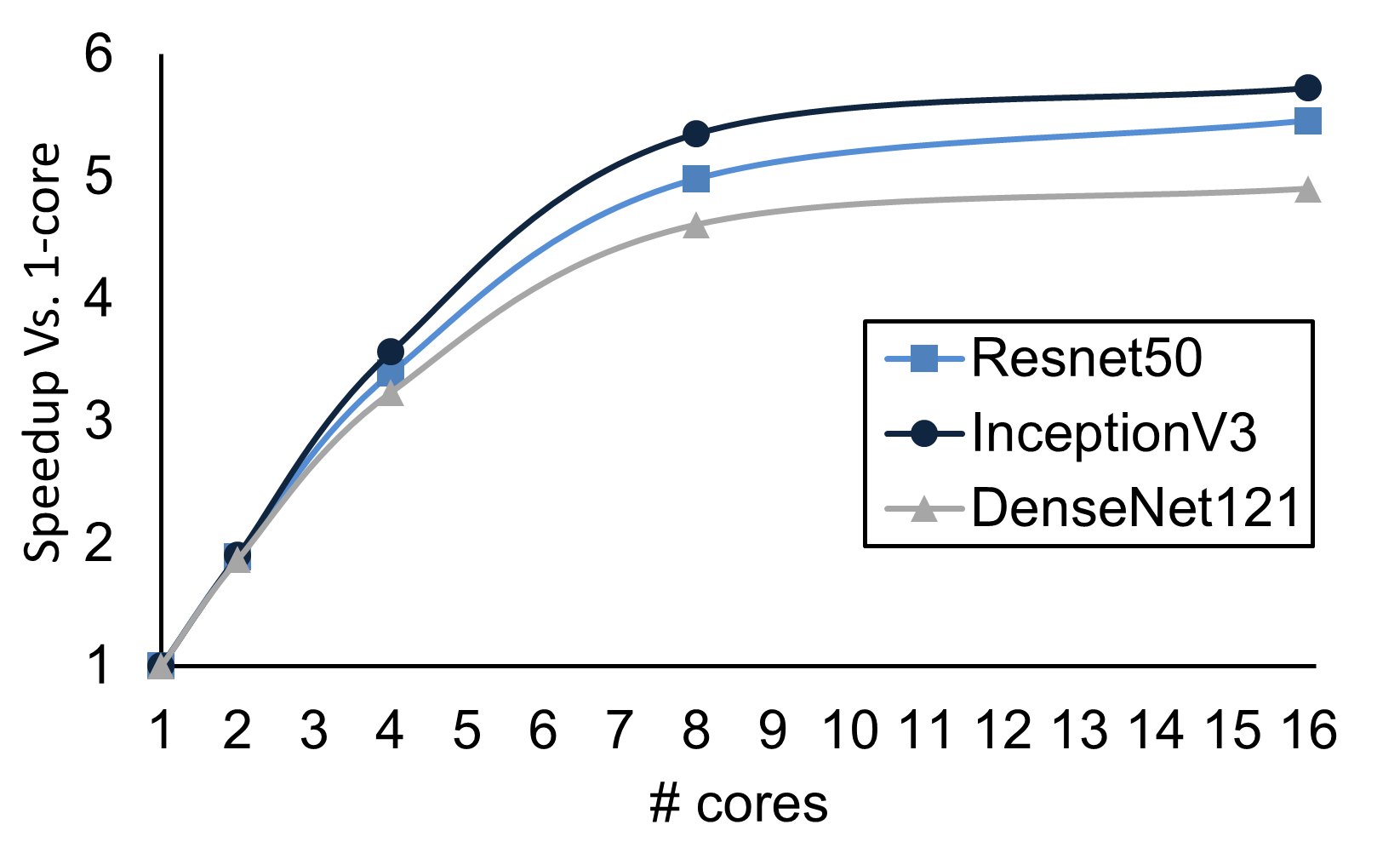}
\caption{The speedup of the proposed approach (`Proposed(8,4)') for CNNs with 1:4 sparsity when executed in a multicore environment, as the number of cores increases. Execution time is normalized to the one achieved by a single core.}
\label{f:multicore}
\end{figure}

Besides the scaling of the hardware vector length, additional parallelism can be extracted by utilizing a \textit{multicore} setup. In such environment, every CNN layer would be executed across multiple cores. The data assigned to each core corresponds to a vertical segment of VL columns of matrix $B$ and the whole structured-sparse matrix $A$. This partitioning is shown in Section~\ref{s:tiling-vindexmac}.

Each core consists of the scalar unit and the decoupled vector unit, as presented in Section~\ref{ss:exp-setup}. All cores share access to the DDR main memory~\cite{gem5_ddr_models} via a cache-coherent interconnect (specifically, the ARM AMBA Coherent Hub Interface (CHI)~\cite{CHI}) that transfers data at the cache-line level.

The speedup achieved when executing the examined CNNs in a multicore setup, as compared to a single-core implementation, is shown in Fig.~\ref{f:multicore}. We tested various multicore CPU sizes, doubling the number of cores in each experiment. The obtained results reveal that performance improves significantly with the number of cores. The reaped speedup gains continue up to 8 cores. Beyond 8 cores, the performance saturates. This is attributed to the limitations of the memory sub-system. As more cores request memory access simultaneously, the memory bandwidth is insufficient to handle the escalating demand; i.e., the bandwidth inevitably saturates. This saturation causes contention among cores for memory resources, which limits the algorithm's scalability and prevents further speedup.

\begin{figure}[t]
\centering
\includegraphics[width=0.7\columnwidth]{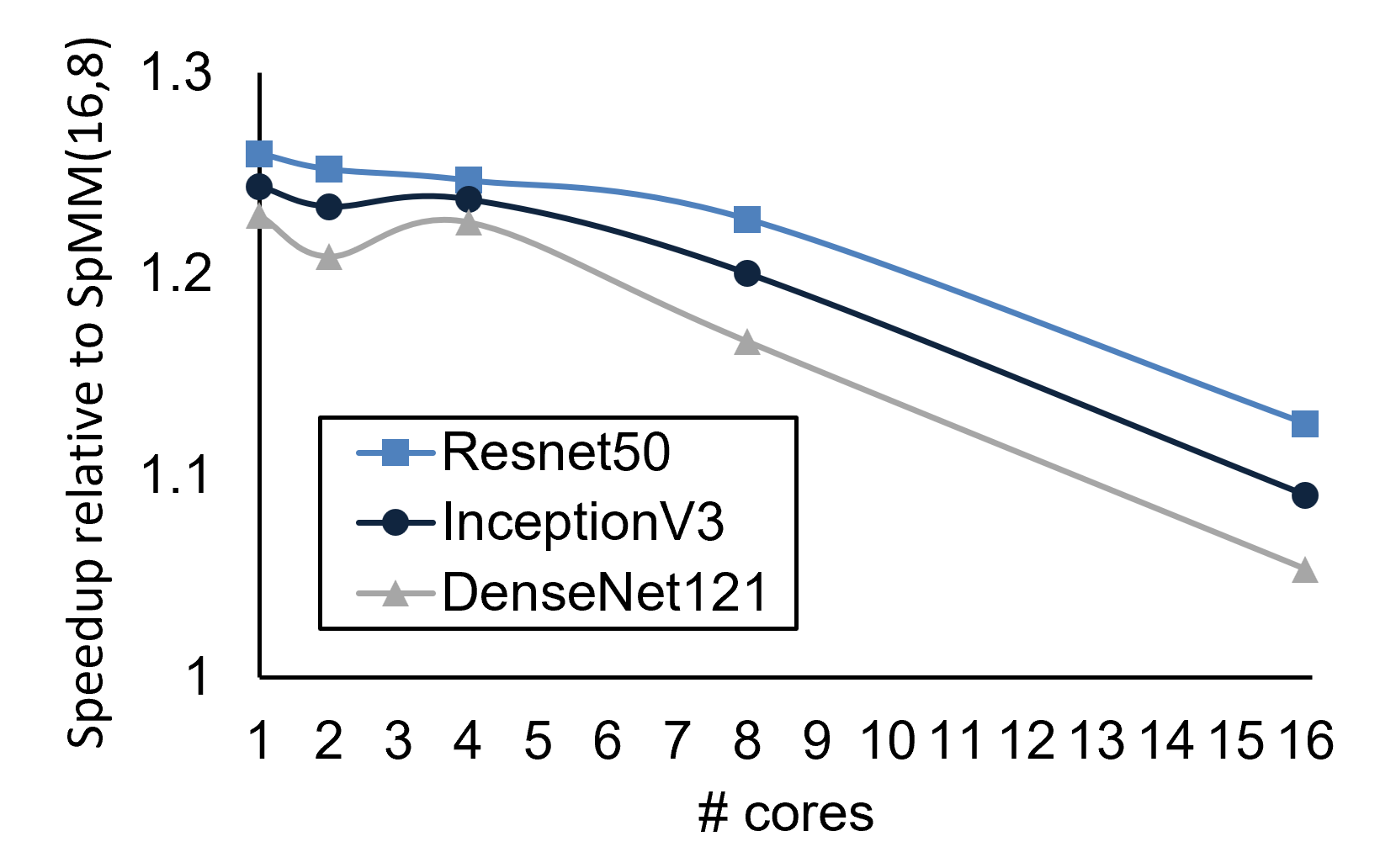}
\caption{The speedup of the proposed approach (`Proposed(8,4)') relative to a baseline multicore setup with unmodified ISA (`SpMM(16,8)') with increasing numbers of cores, for CNNs with 1:4 sparsity.}
\label{f:relative_speedup_multicore}
\end{figure}

This memory bandwidth saturation is also evident in Fig.~\ref{f:relative_speedup_multicore}, which compares the speedup of the proposed approach (Proposed(8,4)) relative to a baseline multicore setup with unmodified ISA (SpMM(16,8)) with increasing numbers of cores. As can be observed, the reaped speedup starts to decrease beyond 8 cores and tends toward 1 (i.e., performance parity) as the cores increase to 16 and beyond. In other words, the proposed approach is most effective in systems where the bottleneck is \textit{computation} and not communication, as is typically the case in multicore setups with \textit{many} cores. The key takeaway is that the proposed new {\tt vindexmac} instruction allows for the use of more power-efficient designs with fewer cores by facilitating better memory bandwidth utilization per core.

\section{Related Work}
\label{s:Related work}

The hardware acceleration of efficient matrix multiplication -- also involving sparse matrices -- has been the focus of extensive research activity within the computer architecture community. Optimizations to sparse matrix multiplication predominantly aim to reduce the number of unnecessary operations involving zero-value elements. As a first step, data storing formats like Compressed Sparse Row (CSR)~\cite{esc} and Compressed Sparse Column (CSC)~\cite{upc-spgemm} help reduce the memory footprint and data movement costs.

Vector processors~\cite{tarantula, xuantie, cornell-vector, ara, riscv-vector, vitruvius} have been utilized extensively to accelerate \textit{dense} matrix multiplication due to their Single Instruction Multiple Data (SIMD) capabilities. Within the context of \textit{sparse} data, various vectorized matrix multiplication implementations have also been proposed~\cite{spa, esc, generic-spgemm, upc-spgemm}. The use of custom vector instructions has also been proposed as a means to accelerate sparse computations~\cite{islped}. These new RISC-V vector data transfer instructions integrate/fuse the memory address generation operation with the standard load operation to reduce energy consumption~\cite{islped}. 

Recent research has primarily focused on \textit{specialized} vector-like hardware architectures to tackle the challenges involved when processing sparse data access patterns, such as those encountered in scientific computing, AI, and data analytics. The Tensor Marshaling Unit (TMU)~\cite{tmu} architecture utilizes a programmable dataflow engine to efficiently handle sparse tensor operations, thereby decoupling data marshaling from computation. A vector-based hardware implementation that aims to improve sparse computations is the Vector Indexed Architecture (VIA)~\cite{via}. VIA leverages a specialized scratchpad memory to improve sparse matrix operation efficiency by optimizing memory traffic and by handling the index-matching process that is central to sparse computations. This approach yields performance gains in sparse matrix-vector multiplication, matrix addition, and matrix multiplication.

Collectively, these architectures represent significant strides in decoupling data access from computation within vector processors. The ultimate goal is to mitigate memory access inefficiencies and enhance parallelism when processing sparse data, which, in turn, can be classified as either unstructured~\cite{rigl}, or structured~\cite{nvidia-block-sparse,learning-n-m}. Alongside the row-wise approach~\cite{matraptor, arm-patent, mm-gp-simd, takayashiki2023new} to matrix multiplication, which aligns well with the capabilities of vector processors, structured sparsity within vector processors presents an intriguing area for further exploration.

Outside the vector processing domain, there has been extensive research pertaining to the general challenges involved with the processing of sparse data. The SpChar framework~\cite{spchar} utilizes decision trees to characterize workload patterns across various sparse kernels, like Sparse Matrix-Vector Multiplication (SpMV), Sparse General Matrix-Matrix Multiplication (SpGEMM), and Sparse Matrix Addition (SpADD) on ARM architectures. This approach provides insights into the influence on performance of key architectural characteristics, such as cache size, memory latency, and branch prediction. Gupta et al.~\cite{dnn} highlight the unique requirements of DNN-based recommendation systems. Unlike traditional DNN workloads, recommendation systems are characterized by irregular memory access patterns -- due to sparse data -- and high memory capacity needs, which are the consequence of utilizing large embedding tables. Finally, FPGA-based accelerators~\cite{fpga} have shown promise in sparse computing, due to their balance of power efficiency and performance, despite potential bandwidth limitations.

\section{Conclusions}
\label{s:conclusions}

Vector processors can efficiently handle the abundant data-level parallelism available in modern ML applications. The scalability of ML models calls for appropriate model pruning that reduces their memory footprint and makes them amenable to applications at the edge. Pruning can benefit various neural network layers, such as fully connected, convolutional, recurrent, and normalization. Furthermore, pruning can benefit attention mechanisms. In this context, we aimed to seamlessly integrate the simplicity of structured sparsity with mainstream vector architectures enhanced with a proposed new instruction.

Initially, we reorganized and optimized the state-of-the-art row-based matrix multiplication algorithm for structured-sparse data for the current definition of the RISC-V ISA. We adopted a hybrid placement for the non-zero data elements and their corresponding column indexes across the vector and the scalar register files. This approach reduced register name dependencies and simplified loop unrolling, thereby significantly improving the runtime. 

Subsequently, this optimized approach was further enhanced with the proposed new {\tt vindexmac} instruction, which operates on local data that is pre-loaded into the vector register file. The use of this new instruction reduces the total number of instructions executed and eliminates unnecessary vector loads. 
Most importantly, the new instruction can be implemented with negligible hardware cost. The evaluation results demonstrate substantial speedups in the execution latency of 
state-of-the-art CNN applications.

\bibliographystyle{IEEEtran}
\bibliography{refs}

\vspace{-1cm}
\begin{IEEEbiography}[{\includegraphics[width=1in,height=1.25in,clip,keepaspectratio]{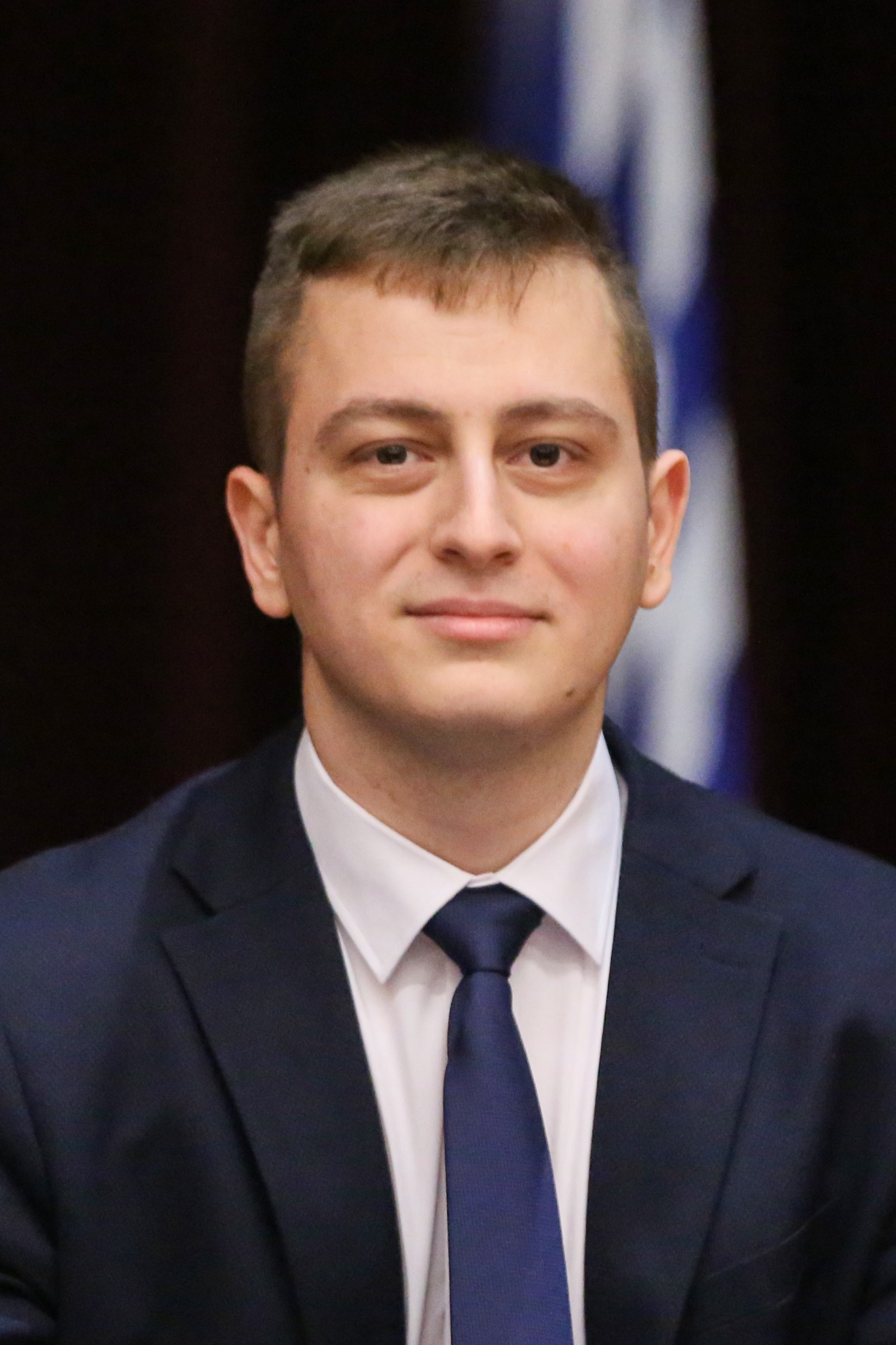}}]{Vasileios Titopoulos} received the Diploma degree in electrical and computer engineering from Democritus University of Thrace, Xanthi, Greece, in 2022, where he is currently pursuing the Ph.D. degree. His research interest includes the design of energy efficient vector processor architectures.
\end{IEEEbiography}
\vspace{-1.3cm}
\begin{IEEEbiography}[{\includegraphics[width=1in,height=1.25in,clip,keepaspectratio]{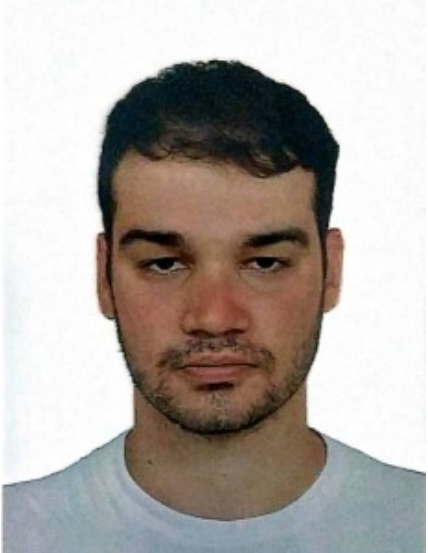}}]{Kosmas Alexandridis} received the Diploma degree in electrical and computer engineering from Democritus University of Thrace, Xanthi, Greece, in 2023, where he is currently pursuing the Ph.D. degree. His research interests include the use of machine learning techniques for verification and design automation of integrated circuits.
\end{IEEEbiography}
\vspace{-1.3cm}
\begin{IEEEbiography}[{\includegraphics[width=1in,height=1.25in,clip,keepaspectratio]{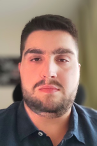}}]{Christodoulos Peltekis} received the Diploma degree in electrical and computer engineering from Democritus University of Thrace, Xanthi, Greece, in 2021, where he is currently pursuing the Ph.D. degree. His research interests include the design of systolic array architectures for machine-learning applications and high-level synthesis design flows. Also, he received the best student paper award at the Artificial Intelligence Circuits and Systems (IEEE AICAS) conference in 2024.
\end{IEEEbiography}
\vspace{-1.3cm}
\begin{IEEEbiography}[{\includegraphics[width=1in,height=1.25in,clip,keepaspectratio]{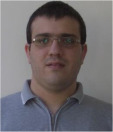}}]{Chrysostomos Nicopoulos} received the B.S. and Ph.D. degrees in electrical engineering with a specialization in computer engineering from Pennsylvania State University, State College, PA, USA, in 2003 and 2007, respectively.

He is currently an Associate Professor with the Department of Electrical and Computer Engineering, University of Cyprus, Nicosia, Cyprus. His current research interests include networks-on-chip, computer architecture, multimany-core microprocessor and computer system design.
\end{IEEEbiography}
\vspace{-1.3cm}
\begin{IEEEbiography}[{\includegraphics[width=1in,height=1.25in,clip,keepaspectratio]{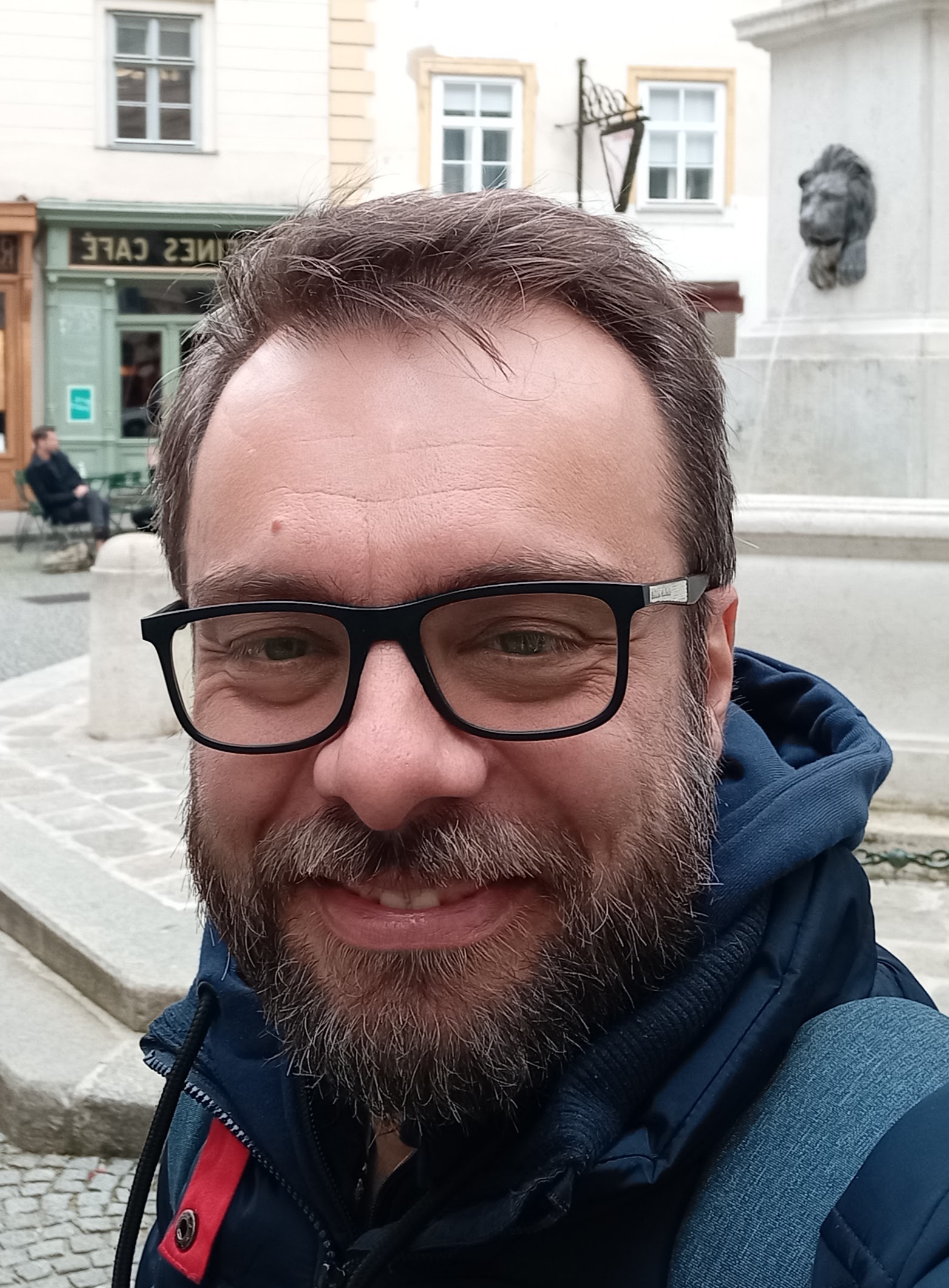}}]{Giorgos Dimitrakopoulos} received the B.S., M.Sc., and Ph.D. degrees in Computer Engineering from the University of Patras, Patras, Greece, in 2001, 2003, and 2007,
respectively. 

He is currently an Associate Professor with the Department of Electrical and Computer Engineering, Democritus University of Thrace, Xanthi, Greece. He is interested in the design of digital integrated circuits, energy-efficient data-parallel accelerators, functional safety architectures, and the use of high-level synthesis for agile chip design.

He received two Best Paper Awards at the Design Automation and Test in Europe (DATE) Conference in 2015 and 2019, respectively. Also, he received the HIPEAC Technology Transfer Award in 2015.
\end{IEEEbiography}

\end{document}